%% file: main.tex
\newif\ifShowComments

\ShowCommentstrue
\documentclass[sigconf,nonacm]{acmart}

\usepackage{xurl}
\usepackage{bm}
\usepackage{xcolor}
\usepackage{graphicx}
\usepackage{pdfpages}
\usepackage[utf8]{inputenc}
\usepackage{amsmath}
\usepackage{amsfonts}
\usepackage{mathtools}
\usepackage{tabularx}
\usepackage{booktabs}
\usepackage{xspace}
\usepackage{boxedminipage}
\usepackage[inline]{enumitem}
\usepackage{amsthm}
\usepackage{multirow}
\usepackage{multicol}
\usepackage{pifont}
\usepackage{makecell}
\usepackage{adjustbox}
\usepackage{hyperref}
\usepackage{algorithm}
\usepackage{thmtools}
\usepackage{color}
\usepackage{xcolor}
\usepackage{stfloats}
\usepackage[noend]{algpseudocode}
\usepackage[capitalize,noabbrev]{cleveref}
\usepackage{subcaption}
\usepackage{float}
\usepackage{soul}
\usepackage{threeparttable}
\usepackage{balance}
\renewcommand\footnotetextcopyrightpermission[1]{} 
\input{setup}

\algdef{SE}[EVENT]{Upon}{EndUpon}[1]{\textbf{upon} #1 \textbf{do}}{\textbf{end upon}}
\algtext*{EndUpon}

\begin{document}
\title{Angelfish: Leader, DAG, or Anywhere in Between}

\author{Qianyu Yu}
\affiliation{
\institution{
The Hong Kong University of Science and Technology (Guangzhou)
}
\city{Guangzhou}
\country{China}
}
\email{qyu100@connect.hkust-gz.edu.cn}

\author{Giuliano Losa}
\affiliation{
\institution{
Stellar Development Foundation
}
\city{San Francisco}
\country{USA}
}
\email{giuliano@stellar.org}

\author{Nibesh Shrestha}
\affiliation{
\institution{
Supra Research
}
\city{Rochester}
\country{USA}
}
\email{n.shrestha@supra.com}

\author{Xuechao Wang}
\authornote{Correspondence: Xuechao Wang}
\affiliation{
\institution{
The Hong Kong University of Science and Technology (Guangzhou)
}
\city{Guangzhou}
\country{China}
}
\email{xuechaowang@hkust-gz.edu.cn}

\renewcommand{\shortauthors}{Qianyu Yu, Giuliano Losa, Nibesh Shrestha, and Xuechao Wang}

\input{abstract}


\maketitle

\input{introduction}

\input{technical_overview}
\input{protocol}
\input{multi-leader}
\input{evaluation}
\input{related_work}
\input{conclusion}
\begin{acks}
Xuechao Wang is supported by the Guangdong Basic and Applied Basic Research Foundation (No. 2025A1515110040), the Guangzhou-HKUST(GZ) Joint Funding Program (No. 2025A03J3882), the Guangzhou Municipal Science and Technology Project (No. 2025A04J4168), and a gift from the Stellar Development Foundation.
We thank the anonymous CCS reviewers for their valuable comments.
\end{acks}

\input{Open_Science}

\bibliographystyle{plain}
\bibliography{main}
\appendix
\input{RBC_denotion}
\input{security_analysis}
\input{multi-leader_security_analysis}
\input{multi_leader_helper}
\input{specs}
\end{document}

%% file: setup.tex
\ifShowComments
\newcommand{\nibesh}[1]{{\footnotesize{\color{orange} [Nibesh:  #1]}}}
\newcommand{\QY}[1]{{\footnotesize{\color{magenta} [Qianyu: #1]}}}
\newcommand{\giuliano}[1]{{\footnotesize{\color{violet} [Giuliano: #1]}}}
\newcommand{\xuechao}[1]{{\footnotesize{\color{red} [Xuechao: #1]}}}
\else
\newcommand{\kartik}[1]{}
\newcommand{\aniket}[1]{}
\newcommand{\nibesh}[1]{}
\newcommand{\QY}[1]{}
\newcommand{\giuliano}[1]{}
\newcommand{\xuechao}[1]{}
\fi

\newcommand{\ignore}[1]{}

\newcommand{\nodes}{\mathcal{P}}
\newcommand{\node}[1]{\ensuremath{P_{#1}}\xspace}

\newcommand{\tuple}[1]{\langle #1 \rangle}

\newcommand{\True}{\mathsf{true}}
\newcommand{\False}{\mathsf{false}}

\algrenewcommand\textproc{}
\newcommand{\N}{\mathbb{N}}

\newcommand{\BigS}{\mathcal{S}}
\newcommand{\BigW}{\mathcal{W}}
\newcommand{\VC}{\mathcal{VC}}

\definecolor{yescolor}{HTML}{026378}
\def\yes{\textcolor{yescolor}{\checkmark}}
\def\no{\textcolor{red}{\ding{55}}}

\newcommand{\name}{Angelfish\xspace}

\makeatletter
\renewcommand{\ALG@name}{}
\makeatother

\algrenewcommand\textproc{}
\algdef{SE}[EVENT]{Event}{EndEvent}[1]{\textbf{upon}\ #1\ \algorithmicdo}{\algorithmicend\ \textbf{event}}%
\algtext*{EndEvent}

\newcommand{\Timeout}{\texttt{timeout}}
\newcommand{\NoVote}{\texttt{no\text{-}vote}}
\newcommand{\TC}{\mathcal{TC}}
\newcommand{\NVC}{\mathcal{NVC}}
\newcommand{\ML}{\mathcal{ML}}
\newcommand{\CLS}{\mathcal{CLS}}
\newcommand{\CMV}{\mathcal{CMV}}

\newcommand{\CVC}{\mathcal{CVC}}

\newcommand{\iunderline}[1]{\noindent\underline{#1}}
\newcommand{\mypara}[1]{\medskip\noindent\textbf{#1}}

\newtheorem{claim}{Claim}
\newtheorem{lemma}{Lemma}
\newtheorem{fact}{Fact}
\newtheorem{property}{Property}
\newtheorem{corollary}{Corollary}

\theoremstyle{definition}
\newtheorem{definition}{Definition}[section]

\newcommand{\sig}[1]{\langle #1 \rangle}

\def\k{\ensuremath{\kappa}\xspace}

%% file: abstract.tex
\begin{abstract}
    To maximize performance, many modern blockchain systems rely on eventually synchronous, Byzantine fault-tolerant (BFT) consensus protocols. Two protocol designs have emerged in this space: protocols that minimize latency using a leader that drives both data dissemination and consensus, and protocols that maximize throughput using a separate, asynchronous data dissemination layer. Recent protocols such as Partially Synchronous Bullshark and Sailfish combine elements of both approaches by using a DAG to enable parallel data dissemination and a leader that paces DAG formation. This improves latency while achieving state-of-the-art throughput. Yet leader-based protocols still achieve better latency under moderate loads, which are common in practice.

    We present \name, a hybrid protocol that adapts smoothly across this design space, from leader-based to Sailfish-like DAG-based consensus. \name lets a dynamically adjusted subset of parties use best-effort broadcast to issue lightweight votes instead of reliably broadcasting costlier DAG vertices. This reduces communication, helps lagging nodes catch up, and lowers latency in practice compared to prior DAG-based protocols. Our empirical evaluation shows that \name attains state-of-the-art peak throughput while significantly lowering latency under moderate load, delivering the best of both worlds.
\end{abstract}

%% file: introduction.tex
\section{Introduction}

State machine replication (SMR)~\cite{lamport_time_1978} is a foundational building block in distributed computing, enabling a set of parties to jointly simulate a single, reliable state machine despite failures and malicious attacks.
In particular, blockchain systems use the SMR approach to provide open, permissionless access to shared virtual machines that users can program using smart contracts.

Applying the SMR approach in a permissionless environment is challenging because the parties participating in an SMR protocol cannot be trusted and may behave maliciously.
A core building block for implementing SMR in such adversarial environments is a Byzantine fault-tolerant (BFT) consensus protocol~\cite{lamport_byzantine_1982}.
BFT consensus protocols allow parties to agree on the sequence of commands, called transactions in a blockchain context, that the state machine must execute, and they are resilient to malicious attacks by some fraction of the parties executing the protocol.

As the consensus protocol is in the critical path of user requests to an SMR system, its throughput and latency are crucial.
A pragmatic approach to devising high-performance consensus protocols is to use the partial synchrony model~\cite{dwork1988consensus}.
This model formalizes the observation that practical networks are mostly synchronous, making high performance possible, but may suffer from bounded periods during which message delays are unpredictable.
Among partially synchronous protocols, two competing approaches have emerged.
The key difference lies in how much the two main tasks of a consensus protocol, namely the dissemination of transaction blocks and their ordering (also called consensus), are decoupled.

On the one hand, BFT consensus protocols~\cite{castro1999practical,yin2019hotstuff,buchmanLatestGossipBFT2019} inspired by traditional partially synchronous consensus protocols~\cite{dwork1988consensus,lamport_part-time_1998} rely on a rotating leader that is responsible for both disseminating transactions and proposing an ordering.
This approach, which tightly couples consensus with data dissemination, can achieve optimal theoretical latency (i.e., 3 message delays~\cite{lamportLowerBoundsAsynchronous2003}) and low latency in practice.
Moreover, while the leader's bandwidth constituted a throughput bottleneck in earlier approaches, protocols like DispersedSimplex~\cite{shoup2024sing} use asynchronous information-dispersal schemes (AVID)~\cite{cachin2005asynchronous} that rely on erasure codes to harness the bandwidth of all parties to help a leader disseminate its transaction block.
This holds the promise of eliminating the leader bottleneck while keeping latency low.
However, erasure-code AVID has computational costs, synchronization costs, and data-expansion costs; for example, in DispersedSimplex, erasure coding increases block size by a factor of roughly 3.
These overheads impact both throughput and latency.

On the other hand, some protocols choose to decouple data dissemination from ordering: an asynchronous data-dissemination layer (sometimes called a mempool) reliably stores transaction blocks, and consensus only orders references to stored transaction blocks.
Examples include Jolteon~\cite{gelashvili2022jolteon}, where stored blocks are unstructured; Narwhal-Hotstuff~\cite{danezis2022narwhal}, which arranges blocks in a DAG (which allows committing full causal histories at once); or Autobahn~\cite{giridharan2024autobahn}, which arranges batches in per-node chains to reduce overhead due to references.
These protocols achieve high throughput because all parties can work to produce and store blocks in parallel and without waiting for ordering, and there is no need for erasure coding and its associated overheads.
However, latency suffers, as it becomes the sum of the data-dissemination latency plus the consensus latency.

Recent research strives to close the latency gap between the two approaches.
To improve latency, DAG-based consensus protocols in the Bullshark~\cite{spiegelman2022bullsharkpartially} family (i.e., Bullshark, Shoal~\cite{spiegelman2023shoal}, Shoal++~\cite{arun2025shoal++}, Sailfish~\cite{shrestha2024sailfish}, etc.) avoid using a separate consensus protocol and instead interpret the DAG structure to reach consensus.
While all parties still build the DAG in parallel, as in previous DAG-based protocols, DAG building is not asynchronous\footnote{By the FLP theorem~\cite{fischerImpossibilityDistributedConsensus1983}, in the absence of randomization, interpreting the DAG to reach consensus requires introducing timeouts in the DAG-building process.}: it evolves in rounds led by rotating leaders that set the pace of DAG formation, and DAG formation stalls until a timeout if a round leader fails.
This approach has successfully reduced the good-case latency of DAG-based protocols while keeping their throughput higher than what leader-based protocols can promise, even when using erasure-coded AVID\footnote{Data expansion caused by erasure coding, typically a factor of around 3, means that blocks can be delivered at a throughput that is only a third of the total available bandwidth.}.
In particular, Sailfish and Shoal++ achieve the theoretical best-case latency lower bound (3 message delays) and state-of-the-art throughput.

However, under workloads that do not saturate bandwidth, even the latency of Sailfish, while theoretically optimal, remains higher in practice than the latency that can be achieved by leader-based protocols.
The main reason is that Shoal++ and Sailfish must send more bits per protocol round: they use reliable broadcast to disseminate DAG vertices, which incurs a total communication cost of \(O(n^3)\) bits per round in the good case; optimal-latency leader-based protocols instead incur a communication cost of \(O(n^2)\) bits per round.

In this work, we propose \name, a new hybrid protocol that achieves the best of both worlds by smoothly adapting between leader-based behavior and DAG-based behavior in order to optimize latency for the load experienced by the system.
In \name, parties build a DAG as in previous DAG-based protocols, but, except for the round leader, they can abstain from proposing DAG vertices via reliable broadcast (RBC) and instead only use the cheaper best-effort broadcast (BEB) primitive to send lightweight vote messages.
As in leader-based protocols, these vote messages contain no transaction blocks and no more than one reference.

At one extreme, when only the leader proposes a DAG vertex in each round, \name behaves similarly to a leader-based protocol and achieves low latency; at the other extreme, when every party proposes a DAG vertex in a round, \name behaves similarly to the state-of-the-art Sailfish protocol~\cite{shrestha2024sailfish} and matches its throughput and latency.
\name is able to continuously adapt between those two extremes in order to obtain the best possible latency given the incoming transaction throughput.
This design poses significant algorithmic challenges, but it offers compelling advantages.

\mypara{Advantage 1: Improved commit latency in practice under moderate load.} Theoretically, in the best case, \name achieves the optimal leader commit latency of three message delays~$\delta$, as do Sailfish~\cite{shrestha2024sailfish} and Shoal++~\cite{arun2025shoal++} (see \Cref{tab: comparison} for a comparison of theoretical latency with other major DAG-based protocols).
In practice, however, \name achieves lower latency when a portion of DAG vertices are replaced by BEB votes.
This is because, in geo-distributed settings, there is likely to be significant variance in message delay between parties, and the speed of round progression is limited by tail latency: to reach the quorum threshold, we need to wait for the slowest of the \(n-f\) fastest parties. Thus, switching a fraction of the parties from RBC vertices, which take two message delays, to BEB votes, which take one message delay, is likely to improve tail latency and make round progression faster. In our evaluation, we observe a significant beneficial impact on latency.

Of course, votes contain no transactions and so make no contribution to throughput.
However, even in popular production DAG-based systems such as Sui (which uses the Mysticeti protocol~\cite{babel2023mysticeti}), the network often operates under moderate load, with most proposed vertices containing only a single transaction or none at all~\cite{suiscan}. With \name, a node with few transactions to propose can use best-effort broadcast to send a vote at a cost of \(O(n)\) bits and one message delay instead of using reliable broadcast to send a DAG vertex at a cost of \(O(n^2)\) bits and two message delays. This improves system-wide latency without decreasing throughput.

In~\Cref{sec: evaluation}, we evaluate \name experimentally on a geo-distributed testbed with 50 nodes.
Compared to Sailfish, \name achieves a reduction in latency of at least 40\% (from roughly 1.0\,s to 0.6\,s) in the moderate-throughput regime (30,000 to 100,000 transactions per second (tps), with 512-byte transactions), under a vertex-proposal rate of 40\% (i.e., when 60\% of the parties are casting votes instead of creating DAG vertices).
This is only about three times the 220 ms latency of our lower-bound baseline: the optimal-latency (\(3\delta\)) leader-based protocol Hydrangea~\cite{shresthaHydrangeaOptimisticTwoRound2025}, which we push to the physical limits of the testbed by using empty blocks (and no mempool).
Moreover, our experiments confirm that \name's latency gains come at no cost to throughput, as peak throughput matches Sailfish.

In addition, we introduce Multi-leader \name, which supports multiple leaders within the same round and further reduces commit latency in practice.
Our evaluation shows that, with 20 leaders, Multi-leader \name achieves a latency of roughly 300\,ms at a vertex-proposal rate of 40\% while serving 25,000\,tps; this latency is only 36\% slower than Hydrangea when it orders empty blocks.

\mypara{Advantage 2: Improved communication complexity.}
Votes also incur lower communication complexity than vertices because only a few different votes can be cast in a given round, so we can aggregate them using multi-signatures~\cite{itakura1983public, anoprenko2025dags}.

The fifth column of \Cref{tab: comparison} compares the communication complexity of \name and other DAG-based protocols under the assumption that RBC proceeds optimistically, as in Narwhal~\cite{danezis2022narwhal}, incurring only \(O(\k n^2)\) complexity in the good case.
Compared to other protocols in the table, with \name we can decrease communication complexity as the load on the system decreases: \name achieves a per-round cost of \(O(\k xn^2 + n^3)\), where $\k$ is the security parameter of the cryptographic primitives used, and $x$ is the average number of parties proposing vertices per round (with each vertex therefore containing $x$ references on average).\footnote{The \(n^3\) term corresponds to the broadcast of vote certificates as BLS multi-signatures, which must include a linear-sized list of signers.}
The last column of \Cref{tab: comparison} shows communication complexity assuming bad-case RBC complexity; in this case, the \(O(n^3)\) complexity of votes is hidden by the cost of RBC of vertices.
The per-round communication complexity of \name is $O(\k xn^3)$, while that of other DAG-based protocols reaches $O(\k n^4)$.
This is the case even in Cordial Miners~\cite{keidar2023cordial}, Mysticeti~\cite{babel2023mysticeti}, and Black Marlin~\cite{ignacio2025dag}, where parties disseminate vertices using BEB and, to ensure liveness, each party must forward all the vertices it receives.

\mypara{Advantage 3: Slower nodes can meaningfully participate.}
In DAG-based protocols like Bullshark and Sailfish~\cite{spiegelman2022bullshark, shrestha2024sailfish}, even if a node does not propose a transaction block, it must use reliable broadcast to disseminate a DAG vertex; this incurs a cost in latency and bandwidth that the node has to pay.
In contrast, \name allows a node not to create a DAG vertex at all and instead to use best-effort broadcast to disseminate a vote message.
This allows slower nodes to minimize the work they have to do to keep up with the rest of the system and, in turn, to meaningfully contribute to system security and fault tolerance with their votes instead of falling behind and risking being considered crashed by the rest of the system.
To validate this claim, we perform an experiment on a real-world testbed in which we make some nodes artificially slow by delaying their traffic by a constant amount of time \(t\). The results, which appear in \Cref{sec: evaluation}, show that slow nodes are able to keep up with much higher delays \(t\) than in Sailfish.


\mypara{Roadmap.} We start in~\Cref{Technical Overview} by giving a high-level technical overview of \name and the challenges in its design. We then provide an in-depth presentation in~\Cref{sec: in-depth} and describe Multi-leader \name in~\Cref{sec: multi-leader}.
\Cref{sec: evaluation} presents an empirical evaluation of both protocols, pitting them against two state-of-the-art protocols, Sailfish~\cite{shrestha2024sailfish} and Autobahn~\cite{giridharan2024autobahn}.
Finally, we discuss related work in \Cref{sec: related work}.

Rigorous security analyses of both protocols and a high-level TLA+ specification are included in the full paper~\cite{yu2025angelfish}. The TLC model checker confirms that the specification satisfies its agreement property under small execution bounds.

\input{related_work_tbl}

%% file: related_work_tbl.tex
\begin{table*}[]
    \footnotesize
\centering
    \caption{Comparison of DAG-based BFT protocols after GST.}
    \setlength\tabcolsep{12pt}
    \def\arraystretch{1}
    \begin{center}
    \label{tab: comparison}
    \begin{tabularx}{\textwidth}{l  c c c c c c }
    \toprule
     &
    \multirow{2}{*}{\makecell{\textbf{Certified} \\ \textbf{DAG}}} &
    \multirow{2}{*}{\makecell{\textbf{LV Commit } \\ \textbf{Latency}}} &
    \multirow{2}{*}{\makecell{\textbf{NLV Commit$^{(1)}$ } \\ \textbf{ Latency}}} &
    \multirow{2}{*}{\makecell{\textbf{Multiple} \\ \textbf{Leaders}}} &
    \multirow{2}{*}{\makecell{\textbf{Bits per Round$^{(2)}$} \\ \textbf{in the Good Case}}} &
    \multirow{2}{*}{\makecell{\textbf{Bits per Round$^{(3)}$} \\ \textbf{in the Bad Case}}}
     \\
     \\
    \midrule
    Bullshark 
    {\cite{spiegelman2022bullshark,spiegelman2022bullsharkpartially}} & 
    {$\yes$} &
    {$4\delta$} &
    {$+2\delta$} &
    {\no}&
    {$O(\k n^3)$} &
    $O(\k n^4)$
    \\
    Shoal 
    {\cite{spiegelman2023shoal}} &
    {$\yes$} &
    {$4\delta$} &
    {$+2\delta$} &
    {$\yes$}&
    {$O(\k n^3)$} &
    $O(\k n^4)$
    \\
    Shoal++~\cite{arun2025shoal++} &
    {$\yes$} &
    {$3\delta$} &
    {$+2\delta$} &
    {$\yes$}&
    {$O(\k n^3)$} &
    $O(\k n^4)$
    \\
    Sailfish~\cite{shrestha2024sailfish} &
    {$\yes$} &
    {$3\delta$} &
    {$+2\delta$} &
    {$\yes$}&
    {$O(\k n^3)$} &
    $O(\k n^4)$
    \\
    Cordial Miners~\cite{keidar2023cordial} &
    {\no} &
    $3\delta$ &
    $+3\delta$ &
    {\no}&
    {$O(\k n^4)$} &
    $O(\k n^4)$
    \\
    Mysticeti~\cite{babel2023mysticeti}&
    {\no} &
    $3\delta$ &
    $+3\delta$ &
    {$\yes$}&
    {$O(\k n^4)$} &
    $O(\k n^4)$
    \\
    \textbf{\name} & 
    {$\yes$} &
    $3\delta$ &
    $+2\delta$ &
    {$\yes$}&
    \bm{$O(\ensuremath{\kappa}xn^2+n^3)$} &
    \bm{$O(\ensuremath{\kappa}xn^3)$}
    \\
    \bottomrule
    \end{tabularx}
    \end{center}
    \begin{flushleft}
    \textbf{LV} stands for leader vertex. \textbf{NLV} stands for non-leader vertex. In \name, $x$ denotes the average number of vertices proposed by honest parties per round.
    (1)~This column lists the additional latency to commit non-leader vertices that share a round with the previous leader vertex; the commit latency of these vertices is the maximum latency among non-leader vertices between two leader rounds.
    (2)~This column lists the communication complexity in the good case when the optimistic RBC path succeeds. We use RBC as in Narwhal~\cite{danezis2022narwhal}, whose optimistic case takes $2$ communication steps and incurs $O(\k n^2)$ communication complexity to propagate an $O(\k n)$-sized message, where $\k$ is a security parameter.
    (3)~This column lists the communication complexity in the bad case, in which parties must initiate the fetching of missing data. This still takes $2$ communication steps to obtain availability certificates but incurs $O(\k n^3)$ communication complexity to deliver an $O(\k n)$-sized message.
    \end{flushleft}
    \label{tbl:related_work}
\end{table*}

%% file: technical_overview.tex
\section{Technical Overview}\label{Technical Overview}
In this section, we first review the technical background on DAG-based protocols. We then present the key idea for reducing latency, the challenges involved in realizing it, and, finally, our solution.

\mypara{DAG-based BFT consensus protocols.}
In DAG-based consensus protocols, each party creates and disseminates so-called vertices, each of which contains a transaction block and cryptographic hashes providing immutable references to other vertices.
This process creates a growing directed acyclic graph (DAG) in which each vertex determines the whole subgraph reachable from it (called its causal history).
Each party then forms its local, partial view of the DAG as the largest set of received vertices forming a DAG with no missing references.

To allow parties to order the vertices of their local DAGs in a consistent way, the protocol defines special vertices called leader vertices (also called anchors in the literature) which are known in advance.
Then, the protocol enforces through its commit rule that any two committed leader vertices (even if committed by two different parties) are connected (by a directed path) in the DAG.
This ensures that, by deterministically ordering the causal history of successive leader vertices, up to the latest committed leader vertex, parties obtain prefixes of the same global total ordering of the DAG.

Different protocols follow this blueprint in different ways, but most protocols build the DAG in rounds where each party creates one vertex per round, including references to at least \(n-f\) vertices from the previous round (where \(n\) is the number of parties and \(f<\frac{n}{3}\) is the maximum number of faulty parties).
This creates a so-called layered DAG, allowing the use of quorum intersection arguments to devise commit rules that interpret the DAG structure to commit leader vertices.

In this work, we aim to improve upon layered-DAG protocols in the Bullshark family, and most notably Sailfish, which achieve theoretically optimal best-case latency.
Such protocols create so-called certified DAGs, in which vertices are considered part of the DAG only after their payload is certified as available.
This ensures that parties can progress from one round to the next without blocking on data retrieval, as long as they obtain an availability certificate.
In practice, this has been shown to be important for robust performance under network instability~\cite{arun2025shoal++}.

\mypara{The latency issue and key idea.}
In certified DAGs like Sailfish~\cite{shrestha2024sailfish} and Bullshark~\cite{spiegelman2022bullshark}, each party must reliably broadcast large vertices (containing transactions and at least $n-f$ references) and wait until it has delivered at least $n-f$ vertices before it can support a leader vertex.
This is a major source of latency compared to classic leader-based protocols.

\name reduces this overhead by allowing non-leader parties to avoid creating DAG vertices and instead cast lightweight votes that either reference only the leader vertex or contain no references at all and that are disseminated using best-effort broadcast.

Since there may be fewer than \(n-f\) vertices created per round, \name dispenses with the requirement that DAG vertices must contain \(n-f\) references to vertices from the previous round.
And, because some parties may use votes instead of creating DAG vertices, \name uses a leader commit rule that counts both votes and DAG vertices: a leader vertex is committed when the number of votes for it plus the number of DAG edges pointing to it from the next round is at least~\(n-f\).

\mypara{The key technical challenge.}
In a protocol like Bullshark or Sailfish, na\"ively removing the requirement that each vertex point to $n-f$ vertices from the previous round would make the protocol unsafe.
To understand why, let us briefly use Sailfish as an example.
In Sailfish, in each round, a rotating leader creates a so-called leader vertex that may possibly be committed.
A leader vertex is committed when $n-f$ parties create vertices in the next round that reference it. Moreover, the leader vertex of the next round must reference it unless there is proof that it cannot be committed. Together with the conditions that each vertex references at least $n-f$ vertices from the previous round and that $n>3f$, a standard quorum-intersection argument ensures that any two committed leader vertices are connected by a directed path, even if there are rounds with no committed leader vertex in between. In turn, this connectivity invariant ensures the safety of the protocol's DAG-linearization rule.

By allowing parties to use lightweight votes that may only contain one reference or no references to previous vertices, we cannot invoke the quorum-intersection argument and, if there are rounds where no leader vertex is committed, we lose the guarantee that committed leader vertices are connected.
Finally, without guaranteed paths between committed leader vertices, locally linearized DAGs may fork.

To overcome this challenge, \name uses special leader edges in the DAG, i.e., direct links between leader vertices that may be more than one round apart, together with new timeout-certificate requirements to guarantee safety across rounds.
Getting this right is the main algorithmic contribution of the paper.
The next paragraphs give an overview of the techniques we employ.

\mypara{Ensuring paths between leader vertices.}
Like most DAG-based protocols, \name uses a rotating round leader, and it requires the round leader to create a vertex (and not send a vote instead).
To ensure that any two committed leader vertices are connected, if a round-$r$ leader vertex $v_r$ does not reference the round-$(r-1)$ leader vertex, it must instead include a leader edge to some prior leader vertex $v_{r'}$ from a round $r' < r-1$, along with $\TC$s ($\Timeout$ certificates) for rounds from $r'+1$ to $r-1$ showing that enough honest parties did not vote for any leader vertex between $v_r$ and $v_{r'}$.
This ensures that the skipped leaders cannot be committed; hence, it is safe for $v_r$ to connect to $v_{r'}$ while bypassing the intermediate leader vertices.
Thanks to reliable broadcast for vertices, in each round we are guaranteed to eventually deliver the leader vertex or receive a timeout certificate, which ensures that leaders can make progress.

\mypara{Pacing round progression.}
Parties should increment their round upon hearing from \(n-f\) parties in the current round (provided they have the leader vertex), whether through votes or vertices, since they cannot wait for more when up to \(f\) parties may be faulty.
A consequence of this design is that some DAG vertices may be delivered too late compared to votes, which can proceed faster, and thus vertices in the next round may not reference them.
As in many DAG protocols (e.g., DAG-Rider~\cite{keidarAllYouNeed2021a}), we use special weak edges to ensure those vertices eventually get included in the DAG.
Additionally, because we do not require \(n-f\) references per vertex or vote, lagging nodes can catch up upon receiving only \(f+1\) messages from a round higher than their next round, versus \(n-f\) in most other DAG-based protocols.
The lightweight jump rule is the mechanism behind Advantage~3.

\mypara{Reducing communication with vote certificates.}
To avoid the need to forward all votes to promote synchronized round progression (since votes do not use RBC), \name relies on parties that complete a round to disseminate certificates summarizing the number of votes they have received. We use multi-signatures to keep bit complexity low. This must be done carefully because there are different possible votes in each round.
This directly realizes Advantage~2: as fewer parties propose vertices, more parties send one of a small number of votes, and these votes can be represented compactly by certificates.

\mypara{The cherry on top: multiple leaders per round.}
To further reduce average latency, \name supports multiple leaders within a single round.
In each round, in addition to the existing main leader, we use a predetermined list of secondary leaders. We require a main leader that has a leader edge to another main leader in a round \(r\) to include, for each secondary leader of \(r\), either an edge or a certificate proving that the secondary leader cannot commit.
Secondary leaders that did not time out are then linearized in order and immediately following the main leader.
There are multiple possible strategies to handle secondary leaders that may not commit, depending on how much waiting for secondary leaders is acceptable.
To balance waiting and including more secondary leaders, we choose to implicitly skip any secondary leader after the first one that fails to commit.

%% file: protocol.tex
\section{\name in Depth}\label{sec: in-depth}
\subsection{Preliminaries}
\label{sec:preliminaries}
\input{preliminaries}


\subsection{Protocol Description}\label{sec:protocol-description}
\mypara{Round-based execution.} Our protocol advances through a sequence of consecutively numbered \emph{rounds}, starting from $1$. In each round $r$, a designated leader, denoted by $L_r$, is selected using a pseudorandom assignment based on the round number.

\input{data_structures_fig}

\input{dag_construction_fig}

\mypara{Basic data structures.} At a high level, the communication among parties is structured as a combination of a DAG and vote messages. In each round, a leader proposes a single vertex containing a block of transactions along with references to vertices proposed in previous rounds, including a distinguished reference to a leader vertex of a previous round. Meanwhile, in each round, a non-leader party either proposes a vertex---if it has transactions to include---or sends a vote. A vertex includes a block of transactions and references to earlier vertices, while a vote references the previous round’s leader vertex or contains no reference. The references in vertices serve as the edges in the DAG. All proposed vertices are propagated using RBC to ensure non-equivocation and guarantee that all honest parties eventually deliver them. The vote messages are propagated using best-effort broadcast (BEB).

\Cref{fig:data_structures} presents the basic data structures and utilities of \name. Each party maintains a local copy of the DAG denoted $DAG_i$. Although honest parties may initially see different local DAGs, the reliable broadcast of vertices ensures that they will eventually converge to a consistent view. Each vertex is uniquely identified by a round number and a sender (source). When $\node{i}$ delivers a vertex from round $r$, it adds the vertex to $DAG_i[r]$, which can contain up to $n$ vertices.

A vertex proposed by $L_r$ is termed the round-$r$ leader vertex, while all other round-$r$ vertices are classified as non-leader vertices. Each non-leader vertex $v$ contains two types of outgoing edges: strong edges and weak edges. In addition to strong edges and weak edges, a leader vertex also contains a leader edge. The strong edges of a vertex $v$ in round $r$ connect to vertices from round $r - 1$, while the weak edges link to at most $n-f$ vertices from earlier rounds $< r - 1$ for which no other path from $v$ exists. A path is a sequence of strong, weak, or leader edges that connects two vertices. For a round-$r$ leader vertex, a leader edge references a leader vertex from some previous round. A leader path is a path from a leader vertex $v_k$ to another leader vertex $v_\ell$ such that all intermediate vertices along the path are leader vertices, and each edge in the path is either a leader edge or a strong edge. Each vertex $v$ also contains $v.tc$ that stores a set of $\Timeout$ certificates (each $\TC$ consisting of a quorum of $\Timeout$ messages in a round). This information helps determine how many vertices a party should wait for in the next round.

In addition to vertex $v$, \name introduces a separate vote structure $vt$. A vote includes a strong edge only when it supports the leader vertex from the previous round; otherwise, it carries no reference.

\mypara{DAG construction protocol.} The DAG construction protocol is outlined in~\Cref{fig:dag_construction}. In each round $r$, a leader proposes a vertex, and a non-leader party $P_i$ proposes either a vertex $v$ or a vote message $vt$, depending on whether it has transactions to propose. To propose a vertex or a vote in round $r$, each party $\node{i}$ waits until it has delivered $n_a$ round-$(r-1)$ vertices and received $n_b$ votes for round $r-1$, such that $n_a+n_b\ge n-f$, with the delivered vertices including the round-$(r-1)$ leader vertex, unless a timeout occurs in round $r-1$ (see Line~\ref{line: advance round}). Here, votes are counted jointly from two sources: (i) received vote messages and (ii) signatures extracted from valid vote certificates. We use $\VC$ to denote a vote certificate that aggregates vote messages. After satisfying the round-entry condition, if votes were used to enter the round, \(\node{i}\) aggregates signatures from the corresponding vote messages and vote certificates to construct \(\VC_{r}^i\) and then broadcasts it.

\begin{figure}[t]
    \centering
    \includegraphics[height=4.5cm]{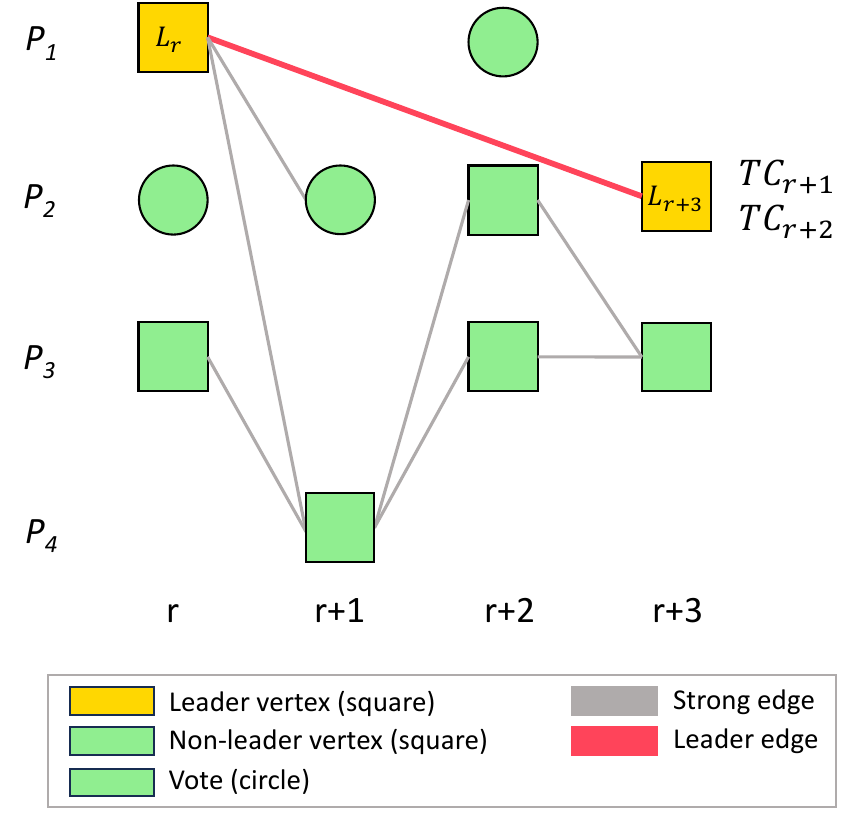}
    \caption{The set of local vertices and votes observed by $P_2$. $L_{r+3}$ did not deliver the round-$(r+1)$ and round-$(r+2)$ leader vertices; therefore, $L_{r+3}$'s vertex must contain $\TC_{r+1}$ and $\TC_{r+2}$.}
    \label{fig:angelfish-illustration}
\end{figure}

After $\node{i}$ successfully advances to round $r$ by delivering the round-$(r-1)$ leader vertex along with at least $n-f$ vertices and votes, if $\node{i}$ intends to send a vote message, it immediately broadcasts $\tuple{vt, r}_i$. Notably, referencing the round-$(r-1)$ leader vertex functions as a ``vote'' for it. These votes are subsequently used to commit the leader vertex. Therefore, waiting for the leader vertex until a timeout ensures that honest parties contribute votes toward it, helping to commit the leader vertex with minimal latency when the leader is honest (after GST).

If an honest party $\node{i}$ does not receive the round-$(r-1)$ leader vertex when its timer expires, it multicasts $\tuple{\Timeout, r-1}$ to all parties (see Line~\ref{line: timeout}). Upon receiving $n-f$ round-$(r-1)$ $\Timeout$ messages (denoted by $\TC_{r-1}$), $P_i$ can enter round $r$ if it has delivered $n_a$ round-$(r-1)$ vertices and received $n_b$ votes such that the total $n_a+n_b$ is at least $n-f$ (see Line~\ref{line: advance round}). It also multicasts $\TC_{r-1}$ so that all other parties will see $\TC_{r-1}$. In \name, we require that if a party sends a round-$(r-1)$ $\Timeout$ message, the round-$r$ vertex or vote it sends will not include a strong edge to the round-$(r-1)$ leader vertex, which serves as a ``no-vote'' for that leader vertex.

We place an additional constraint on the leader vertex. A round-$r$ leader vertex needs to either (1) have a strong edge to the round-$(r-1)$ leader vertex or (2) include a leader edge referencing a leader vertex from some earlier round $r' < r - 1$ and contain a $\Timeout$ certificate $\TC_{r''}$ for each round $r''$ such that $r'+1 \leq r'' \leq r-1$. Each $\TC_{r''}$ serves as evidence that a quorum of parties did not ``vote'' for the round-$r''$ leader vertex. Hence, the round-$r''$ leader vertex cannot be directly committed, and it is safe to lack a leader path to that leader vertex. An illustration is provided in~\Cref{fig:angelfish-illustration}.

Upon delivering a round-$r$ vertex $v$, if $v$ is a leader vertex, each party $P_i$ verifies its validity by invoking is\_valid$(v)$. The function is\_valid$(v)$ returns true if $v$ contains a leader edge referencing a round-$r'$ leader vertex and a $\Timeout$ certificate $\TC_{r''}$ for each round $r''$ such that $r'+1 \leq r'' \leq r-1$. Only valid vertices \(v\) are added to $DAG_i[r]$ via try\_add\_to\_dag($v$). For non-leader vertices, this validity check is not required; they are passed directly to try\_add\_to\_dag($v$), which succeeds once $\node{i}$ has delivered all vertices that establish a path from $v$ in $DAG_i$. If try\_add\_to\_dag($v$) fails, the vertex is placed in a \textit{buffer} for a later retry. Additionally, when try\_add\_to\_dag($v$) succeeds, any buffered vertices are retried for inclusion in $DAG_i$.
In parallel, upon receiving a vote message $\tuple{vt,r}_*$, the party $P_i$ gathers its signature into an array. Upon receiving a vote certificate, $P_i$ first verifies the signatures and then stores them in the corresponding array. Once $P_i$ enters round $r$ using votes, it aggregates the signatures from the array to form $\VC_{r}^i$, which it then broadcasts so that other parties can also receive these signatures. Moreover, since vertices are broadcast via RBC, once a vertex is delivered by an honest party, all other parties will eventually deliver it as well. Therefore, once an honest party has collected $n-f$ votes and vertices, all other parties will eventually receive enough votes and vertices to enter the next round.

\input{jump_round_fig}

\mypara{Jumping rounds.} In addition to advancing rounds sequentially, our protocol allows honest parties in round $r' < r-1$ to ``jump'' directly to round $r$. This occurs when they either observe at least $f+1$ round-$r$ vertices and votes in total, including the round-$r$ leader vertex, or receive a $\TC_{r}$ (see \Cref{line: jumping round} in \Cref{fig:jumping round}). We require only $f+1$ vertices and votes, rather than $n-f$, because this already ensures that at least one honest party has advanced to round $r$. 
If the lagging party is $L_r$, then in addition to collecting $f+1$ round-$r$ vertices and votes, it must also wait until it has delivered either the round-$(r-1)$ leader vertex or some earlier leader vertex (say round $r'$), together with the corresponding $\TC$s for rounds between $r'$ and $r-1$, before proposing the round-$r$ leader vertex. When jumping from round $r'$ to round $r$, parties neither propose vertices nor send votes for the skipped rounds. This design allows slow parties to catch up more efficiently.

\name has a better jumping mechanism than Sailfish~\cite{shrestha2024sailfish}. Sailfish requires delivering the leader vertex (or receiving a $\TC$) together with $n-f$ vertices to jump rounds, whereas \name requires only the leader vertex (or $\TC$) plus $f+1$ vertices and votes. This makes slow parties in \name more likely to catch up than in Sailfish. Moreover, when a slow party enters a new round without downloading transactions, it can send a vote message instead of proposing an empty block, thereby reducing communication overhead.

\mypara{Committing and ordering the DAG.} In our protocol, only leader vertices are committed, while non-leader vertices are ordered in a deterministic manner as part of the causal history of a leader vertex. This ordering occurs when the leader vertex is (directly or indirectly) committed, as defined in the order\_vertices function.

\input{commit_rule_fig}

The commit rule is presented in~\Cref{fig:commit-rule}. An honest party $P_i$ commits a round-$r$ leader vertex once it observes $n-f$ vertices and votes in total, including both (i) first RBC messages from round-$(r{+}1)$ vertices that reference the round-$r$ leader vertex, and (ii) votes, including both vote messages that support $v_\ell$ and the commit vote certificate \(\CVC_{r+1}\), which is a collection of signatures from vote messages supporting $v_\ell$. $\node{i}$ does not need to wait for the delivery of round-$(r+1)$ vertices. This is because, when the sender of the RBC is honest, the first observed value (i.e., the first message of the RBC) is the same value that will ultimately be delivered. Among the $n-f$ round-$(r+1)$ vertices and votes, at least $f+1$ originate from honest parties and will eventually be delivered, ensuring that the delivered value matches the first received value (from the first RBC message). Furthermore, since at least $f+1$ honest parties send round-$(r+1)$ vertices and votes with a strong edge to the round-$r$ leader vertex, it is impossible to form a $\TC_{r}$. Consequently, the round $r' > r$ leader vertex (if it exists) will necessarily have a leader path to the round-$r$ leader vertex, ensuring the safety of the commit. Similar to propose vote certificates, if commitment is achieved using votes, then $P_i$ aggregates signatures from the relevant vote messages and commit vote certificates to generate its own commit vote certificate, and then broadcasts it to help other parties commit.

When directly committing a round-$r$ vertex $v_k$ in round $r$, $\node{i}$ traces back through earlier rounds to \textit{indirectly} commit leader vertices $v_m$ if a leader path exists from $v_k$ to $v_m$. This process continues until $P_i$ encounters a round $r' < r$ where it has already directly committed a leader vertex. Our protocol ensures that once any honest party directly commits a round-$r$ leader vertex $v_k$, all leader vertices of subsequent rounds $r' > r$ will have a leader path to $v_k$. This guarantees that $v_k$ will eventually be committed by all honest parties.

\mypara{Remark on timeout parameter $\tau$.} The timeout parameter $\tau$ must be set sufficiently long to guarantee that, upon entering round $r$, an honest party has enough time to deliver both the round-$r$ leader vertex—if broadcast by an honest leader—and at least $n-f$ round-$r$ vertices and votes before its timer expires. The precise value of $\tau$ depends on the underlying RBC primitive used to disseminate vertices, as well as the specific conditions under which an honest party $\node{i}$ transitions to round $r$. When using Bracha’s RBC~\cite{bracha1987asynchronous}, our protocol requires $\tau = 2\Delta$ if $\node{i}$ enters round $r$ after delivering the leader vertex from round $r - 1$, and $\tau = 5\Delta$ otherwise. We prove in the full paper~\cite{yu2025angelfish} that these timeout bounds are sufficient.

\mypara{Remark on communication complexity.}
We now explain the communication complexity of Angelfish in \Cref{tab: comparison}. In each round, $x$ denotes the average number of parties proposing vertices. Then, in the good case, $n-x$ is the average number of parties sending votes. The per-round complexity of vertices is $O(\k xn^2)$ due to the reliable broadcast of $O(\k n)$-sized vertices. The per-round complexity of votes is $O(\k(n-x)n+(n-x+\k)n^2)$. The first term accounts for multicasting the $n-x$ votes of size $O(\k)$. The second term accounts for multicasting the $n$ vote certificates of size $O(n-x+\k)$, as these certificates use multi-signatures. If we combine the vertex and vote complexities, the overall communication complexity simplifies to $O(\k xn^2+n^3)$.

\mypara{Remark on reducing latency.}
We give a simple latency estimate to illustrate why allowing parties to send votes instead of vertices reduces latency. The latency reduction comes mainly from parties being able to enter rounds faster. Let $\delta_{ik}$ be the message latency from party $P_i$ to party $P_k$. For a finite set $S$, let $\operatorname{min}_{m}(S)$ denote the $m$-th smallest value in $S$.

Consider party $P_j$ receiving a vertex proposed by party $P_i$ using two-round RBC. Since the RBC delivery path goes through a quorum of parties, the delivery latency is approximately
\[
T_{i\to j}
=
\operatorname{min}_{n-f}
\left\{
\delta_{ik}+\delta_{kj}
: P_k\in\mathcal{P}
\right\}.
\]
In Sailfish, to enter round $r$, a party
must wait for $n-f$ round-$(r-1)$ vertices, including the round-$(r-1)$ leader vertex. If the leader is $L_{r-1}$, the round-entry latency for a party \(P_j\) is therefore approximately
\[
\max\left\{
\operatorname{min}_{n-f}
\left\{
T_{i\to j}:P_i\in\mathcal{P}
\right\},
T_{L_{r-1}\to j}
\right\}.
\]
In \name, if the propose\_rate is $\alpha$, then about $\alpha n$ parties propose vertices, so a party waits for roughly $\alpha n-f$ RBC-delivered vertices in addition to the leader vertex; the remaining messages can be votes, which take only one message delay and whose latency is likely negligible compared with RBC latency. Thus the round-entry latency is approximately
\[
\max\left\{
\operatorname{min}_{\alpha n - f}
\left\{
T_{i\to j}:P_i\in\mathcal{P}
\right\},
T_{L_{r-1}\to j}
\right\}.
\]
Comparing the two expressions shows that \name reduces round-entry latency by replacing the need to wait for the $(n-f)$-th fastest RBC-delivered vertex with the need to wait for only the $(\alpha n-f)$-th fastest such vertex. The gain is larger when the propose rate $\alpha$ is smaller.

For example, consider $n=10$ and $f=3$, with $P_1,P_2$ in South Carolina, $P_3,P_4$ in Oregon, $P_5,P_6$ in Belgium, $P_7,P_8$ in Finland, and $P_9,P_{10}$ in Japan. Consider party $P_1$ in South Carolina. For each proposer $P_i$, the two-round RBC delivery time of its vertex at $P_1$ is
\[
T_{i\to 1}
=
\operatorname{min}_{7}
\left\{
\delta_{ik}+\delta_{k1}:P_k\in\mathcal{P}
\right\}.
\]
Using the latency values in~\Cref{tab:ping-latencies}, \Cref{tab:rbc-delivery-p1} shows how each $T_{i\to1}$ is obtained. Each region contains two parties, so each region value appears twice in the multiset $\{\delta_{ik}+\delta_{k1}:P_k\in\mathcal{P}\}$; the displayed $T_{i\to1}$ is the seventh value after sorting these ten terms. We write $x^{\times m}$ for $m$ copies of the value $x$.

\begin{table}[ht]
\caption{Two-round RBC delivery latency at $P_1$.}
\label{tab:rbc-delivery-p1}
\centering
\small
\setlength{\tabcolsep}{1.5pt}
\begin{tabular}{@{}c c c c@{}}
\toprule
Proposer(s) & Sorted 2-round RBC latency (ms) & $T_{i\to1}$ & Rank \\
\midrule
$P_5,P_6$ & $94^{\times4},146^{\times2},200^{\times2},391^{\times2}$ & $\delta_{53}+\delta_{31}\approx 200$ ms & 1--2 \\
$P_7,P_8$ & $116^{\times4},124^{\times2},221^{\times2},416^{\times2}$ & $\delta_{73}+\delta_{31}\approx 221$ ms & 3--4 \\
$P_1,P_2$ & $2^{\times2},129^{\times2},186^{\times2},231^{\times2},325^{\times2}$ & $\delta_{27}+\delta_{71}\approx 231$ ms & 5--6 \\
$P_3,P_4$ & $66^{\times4},228^{\times2},257^{\times2},273^{\times2}$ & $\delta_{39}+\delta_{91}\approx 257$ ms & 7--8 \\
$P_9,P_{10}$ & $155^{\times2},169^{\times4},322^{\times2},365^{\times2}$ & $\delta_{95}+\delta_{51}\approx 322$ ms & 9--10 \\
\bottomrule
\end{tabular}
\end{table}

Suppose the previous-round leader is $P_5$ in Belgium. Then the leader vertex arrives at $P_1$ after $T_{5\to1}=\delta_{53}+\delta_{31}$, so it is not the bottleneck. Since each row corresponds to two proposers, Sailfish waits for the seventh fastest RBC-delivered vertex, which is an Oregon vertex with latency $\delta_{39}+\delta_{91}=257$ ms. In \name with $\alpha=0.6$, only $\alpha n-f=3$ non-leader RBC vertices are needed in addition to the leader vertex and the votes, so \name waits for the third fastest RBC-delivered vertex, which is a Finland vertex with latency $\delta_{73}+\delta_{31}=221$ ms. Thus, in \name \(P_1\) enters the next round 36 ms earlier than in Sailfish.

\mypara{Safety and liveness proof sketch.}
The key safety invariant is that a leader vertex directly committed by an honest party cannot be skipped by later valid leader vertices. The reason is that direct commitment requires a quorum of support, while skipping a leader requires timeout certificates; by quorum intersection and the timeout rule, these two conditions cannot both hold for the same leader. Thus later committed leaders remain connected to earlier committed leaders through leader paths. Since reliable broadcast eventually gives honest parties the same delivered vertices and causal histories, the deterministic \Call{order\_vertices}{} procedure makes all honest parties output the same order.

For liveness, first note that in every round, either the leader vertex is delivered by reliable broadcast or we get a timeout certificate (or both).
Thus, each round, we eventually move to the next round (Line~\ref{line: advance round} in~\Cref{fig:dag_construction}).
Moreover, after GST, any evidence that lets one honest party advance, including vertices, votes, vote certificates, and timeout certificates, reaches all honest parties within a bounded delay. Hence honest parties enter each successive new round within some maximum fixed delay of each other.
Thus, after GST, when an honest party is the leader, proposes a vertex, and uses a sufficiently large timeout parameter, all honest parties enter its round, deliver its vertex, and produce enough next-round support for it to be directly committed.

Thus it remains to determine whether, after GST, honest parties can propose.
The only problematic case is if the leader of a round \(r>1\) timed out on the leader of the previous round \(r-1\), but nevertheless no timeout certificate forms for \(r-1\).
In this case, by the rules of the protocol, the leader of round \(r\) cannot propose.
However, note that, after GST and assuming sufficiently long timeouts, this can only happen if the leader of round \(r-1\) is not honest; otherwise, either the leader of \(r-1\) proposed and its proposal is delivered by honest nodes before they time out, or it does not propose and a timeout certificate forms.
We conclude that, after GST, a commit is guaranteed in every round \(r\) that has an honest leader and such that \(r-1\) also has an honest leader and also started after GST.

The complete proof appears in the full paper~\cite{yu2025angelfish}.

\mypara{TLA+ specification.}
In the full paper~\cite{yu2025angelfish}, we additionally provide a TLA+ specification of \name\footnote{https://doi.org/10.5281/zenodo.21855535}.
This specification describes \name at a high level of abstraction, where vertices, votes, and timeout messages are posted atomically on a global bulletin board.
While unrealistic, this allows us to convey clearly and precisely the high-level algorithmic ideas behind \name and can serve as a formal reference that is free of ambiguities.

Using the TLC model checker, we also exhaustively explored the behaviors of the TLA+ specification for 3 honest nodes configured with \(f=1\) (so, any set of 2 is a quorum) executing 3 rounds of \name.
The model checker successfully terminated in 3 hours and 13 minutes using 32 workers on a 32-vCPU AMD EPYC virtual machine with 64 GB of memory; it generated 1,902,582,078 states, of which 18,938,251 were unique, and found no violations of the type-correctness invariant or the protocol's agreement property.
This gives us confidence that the specification is meaningful and free of basic errors, but it does not imply that we verified the protocol in any strong sense.

%% file: preliminaries.tex
We consider a system $\nodes := \node{1},\ldots,\node{n}$ consisting of $n$ parties, of which up to $f<\frac{n}{3}$ can be Byzantine, meaning they can behave arbitrarily. A party that is not faulty throughout the execution is said to be \emph{honest} and follows the protocol.

We consider the partial synchrony model of Dwork et al.~\cite{dwork1988consensus}. In this model, the network initially operates in an asynchronous state, allowing the adversary to arbitrarily delay messages sent by honest parties. However, after an unknown point in time, known as the \emph{Global Stabilization Time} (GST), the adversary must ensure that all messages from honest parties are delivered to their intended recipients within $\Delta$ time of being sent (where $\Delta$ is known to the parties).
We use \(\delta\) for the actual message delay after GST, noting that $\delta \leq \Delta$. Additionally, we assume the local clocks of the parties have \emph{no clock drift} and \emph{arbitrary clock skew}.

We make use of digital signatures and a public-key infrastructure (PKI) to authenticate messages and prevent spoofing or replay attacks. We use $\tuple{x}_i$ to denote a message $x$ digitally signed by party $P_i$ using its private key.


We assume that each party receives transactions from external clients and periodically groups them into blocks that it submits to a Byzantine atomic broadcast service implemented by Angelfish. This service delivers blocks in the same total order at all parties.
\begin{definition}[Byzantine atomic broadcast~\cite{keidarAllYouNeed2021a,spiegelman2022bullshark}]
\label{dfn:bab}
Each honest party $\node{i} \in \nodes$ can call \Call{a\_bcast$_i$}{$b$} to submit a block $b$. Each party $\node{i}$ may then output \Call{a\_deliver$_i$}{$b,\node{k}$}, where $\node{k}\in \nodes$ represents the block creator. A Byzantine atomic broadcast protocol must satisfy the following properties:
\begin{itemize}[noitemsep,leftmargin=*]\sloppy
    \item[-] \textbf{Agreement.} If an honest party $\node{i}$ outputs \Call{a\_deliver$_i$}{$b, \node{k}$}, then every other honest party $\node{j}$ eventually outputs \Call{a\_deliver$_j$}{$b, \node{k}$}.
    
    \item[-] \textbf{Integrity.} For every block $b$ and party $\node{k} \in \nodes$, an honest party $\node{i}$ outputs \Call{a\_deliver$_i$}{$b,\node{k}$} at most once and, if \(\node{k}\) is honest, only if \(\node{k}\) previously called \Call{a\_bcast$_k$}{$b$}.

    \item[-] \textbf{Validity.} If an honest party $\node{k}$ calls \Call{a\_bcast$_k$}{$b$}, then every honest party eventually outputs \Call{a\_deliver}{$b,\node{k}$}.
    
    \item[-] \textbf{Total order.} If an honest party $\node{i}$ outputs \Call{a\_deliver$_i$}{$b, \node{k}$} before \Call{a\_deliver$_i$} {$b', \node{\ell}$}, then no honest party $\node{j}$ outputs \Call{a\_deliver$_j$}{$b', \node{\ell}$} before \Call{a\_deliver$_j$}{$b, \node{k}$}.
\end{itemize}
\end{definition}


%% file: data_structures_fig.tex
\begin{figure*}[!ht]
    \footnotesize
	\begin{boxedminipage}[t]{\textwidth}
		\textbf{Local variables:}
		\setlist{nolistsep}
		\begin{itemize}[noitemsep]
			\item[] struct vertex $v$: \Comment{The struct of a vertex in the DAG}
			      \begin{itemize}[noitemsep]
			      	\item[] $v.round$ - the round of $v$ in the DAG
			      	\item[] $v.source$ - the party that broadcasts $v$
			      	\item[] $v.block$ - a block of transactions   
                        \item[] $v.strongEdges$ - a set of vertices in $v.round-1$ that represent strong edges
    				\item[] $v.weakEdges$ - a set of vertices in rounds $<$ $v.round-1$ that represent weak edges
                        \item[] $v.leaderEdge$ - a leader vertex in a round $<$ $v.round-1$ that represents a leader edge (for leader vertices)
                        \item[] $v.tc$ - a set of $\Timeout$ certificates for round $\le$ $v.round-1$ (if any)
                        \end{itemize}
			\item[] struct vote $vt$: \Comment{The struct of a vote}
			      \begin{itemize}[noitemsep]
			      	\item[] $vt.round$ - the round of $vt$
                        \item[] $vt.source$ - the party that broadcasts $vt$
                        \item[] $vt.sig$ - signature of $vt$
                        \item[] $vt.strongEdge$ - leader vertex in $vt.round-1$ that the party votes for; otherwise, $\bot$ 
			      \end{itemize}
                \item[] struct vote certificate $\VC$: \Comment{The struct of aggregated BLS signatures from votes}
                \begin{itemize}[noitemsep]
                    \item[] $\VC.sig$ - a set of signatures of $\VC$
                \end{itemize}
                \item[] struct commit vote certificate $\CVC$: \Comment{The struct of aggregated BLS signatures from votes referencing the leader vertex}
                \begin{itemize}[noitemsep]
                    \item[] $\CVC.sig$ - a set of signatures of $\CVC$
                \end{itemize}
                \item[] $\Timeout$ certificate $\TC$ - an array of sets, each containing a quorum of $\Timeout$ messages
                \item[] $DAG_i[]$ - an array of sets of vertices
                \item[] $votes[]$ - an array of sets of votes
                \item[] $sigVote[]$ - an array of sets of signatures of parties that propose $vt$
                \item[] $sigCommit[]$ - an array of sets of signatures of parties that propose $vt$ and vote for the leader vertex in $vt.round-1$
                \item[] $vertexProposers[]$ - an array of sets of proposers that propose vertices
			\item[] $blocksToPropose$ - a queue, initially empty; $\node{i}$ enqueues valid blocks of transactions from clients
                \item[] $leaderStack \gets $ an empty stack
                \item[] $willProposeVertex[]$ - if a party intends to propose a vertex
                
		\end{itemize}

		\begin{algorithmic}[1]
			\Procedure{path}{$v,u$} \Comment{Check whether a path consisting of strong, weak, and leader edges exists in the DAG}
			\State \Return \parbox[t]{\dimexpr\textwidth-\leftmargin-\labelsep-\labelwidth-21pt}{there exist $k \in \N$ and vertices $v_1,\ldots, v_k$ s.t. \\
$v_1 = v$, $v_k = u$, and $\forall j \in [2,..,k]: v_j \in \bigcup_{r\ge 1} DAG_i[r] \land (v_j \in v_{j-1}.weakEdges \cup v_{j-1}.strongEdges  \cup  v_{j-1}.leaderEdge)$\strut} 
			\EndProcedure
			
                \Procedure{leader\_path}{$v,u$} \Comment{Check whether a path consisting of leader and strong edges from leader vertex $v$ to leader vertex $u$ exists in the DAG}
			\State \Return \parbox[t]{313pt}{there exist $k \in \N$ and leader vertices $v_1,\ldots, v_k$ s.t.\ \\
				$v_1 = v$, $v_k = u$, and $\forall j \in [2,..,k]: v_j \in \bigcup_{r\ge 1} DAG_i[r] \land v_j \in  v_{j-1}.leaderEdge \cup v_{j-1}.strongEdges $\strut}  
			\EndProcedure
            \algstore{bkbreak}
        \end{algorithmic}
    
    \vspace{0.4em}
    
   \begin{minipage}{0.48\textwidth}
        \begin{algorithmic}
            \algrestore{bkbreak}
            \Procedure{set\_weak\_edges}{$v, r$} \Comment{Add edges to orphan vertices}
            \State $v.weakEdges \gets \{\}$
            \For{$r' = r-2 \text{ down to } 1$}
            \For{\textbf{every} $u \in DAG_i[r']$ s.t.  $\neg$\Call{path}{$v,u$}}
            \State $v.weakEdges \gets v.weakEdges \cup \{u\}$
            \EndFor
            \EndFor
            \EndProcedure

            \vspace{0.4em}
   
			\Procedure{get\_vertex}{$p, r$}
			\If{$\exists v \in DAG_i[r]$ s.t. $v.source = p$}
			\State \Return $v$
			\EndIf
			\State \Return $\bot$
			\EndProcedure
            
            \vspace{0.4em}
			
            \Procedure{get\_leader\_vertex}{$r$}
			\State \Return \Call{get\_vertex}{$L_r, r$} 
			\EndProcedure

            \vspace{0.4em}
        
            \Procedure{a\_bcast$_i$}{$b,r$}
			\State $blocksToPropose.$enqueue($b$)
			\EndProcedure
            
            \algstore{bkbreak}
            \end{algorithmic}
            \end{minipage}
            \hfill
           \begin{minipage}{0.48\textwidth}
            \begin{algorithmic}
            \algrestore{bkbreak}
        
            \Procedure{broadcast\_vertex}{$r$}
			\State $v \gets$ \Call{create\_new\_vertex}{$r$}
			\State \Call{try\_add\_to\_dag}{$v$}
			\State \Call{r\_bcast$_i$}{$v, r$}
			\EndProcedure
            \vspace{0.4em}
            \Procedure{multicast\_vote}{$r$}
                \State $\tuple{vt,r}_i \gets$ \Call{create\_new\_vote}{$r$}
                \State $votes[r]\gets votes[r]\cup \{vt\}$
                \State multicast $\tuple{vt,r}_i$
            \EndProcedure
             \vspace{0.4em}

            \Procedure{order\_vertices}{$ $}\label{func:order-vertices}
			\While{$\neg leaderStack.$isEmpty()}
			\State $v \gets leaderStack.$pop()
			\State $verticesToDeliver \gets \{v' \in \bigcup_{r>0} DAG_i[r] \,| \, path(v, v')  \land v' \not \in deliveredVertices\}$
			\For{\textbf{every} $v' \in verticesToDeliver$ in some deterministic order}
			\State \textbf{output} \Call{a\_deliver$_i$}{$v'.block, v'.round, v'.source$}
			\State $deliveredVertices \gets deliveredVertices \cup \{v'\}$
            \EndFor
            \EndWhile
			\EndProcedure
			\algstore{bkbreak}
		\end{algorithmic}
        \end{minipage}
  
	\end{boxedminipage}
	\caption{Basic data structures for \name. The utility functions are adapted from~\cite{spiegelman2022bullshark,shrestha2024sailfish}.}
	\label{fig:data_structures}
\end{figure*}

%% file: dag_construction_fig.tex
\begin{figure*}[!ht]
	\footnotesize
	\begin{boxedminipage}[t]{\textwidth}
		\textbf{Local variables:}
		\setlist{nolistsep}
		\begin{itemize}[noitemsep]
			\item[] $round \gets 1;$ $\textit{buffer} \gets \{\}$
		\end{itemize}
		
		   \begin{minipage}{0.48\textwidth}
            \begin{algorithmic}
            \algrestore{bkbreak}
	\Upon {\Call{r\_deliver$_i$}{$v, r, p$}} \label{line: deliver vertex} \label{line: accept}
        \If {$v.source = p \land v.round = r \land$ ($v.source=L_r$ $\land$ is\_valid$(v)$)}
            \If {$\neg$\Call{try\_add\_to\_dag}{$v$}} \label{line: is_valid}
                \State $\textit{buffer} \gets \textit{buffer} \cup \{v\}$
            \Else
                \For {$v' \in \textit{buffer} : v'.round \geq r$}
                    \State \Call{try\_add\_to\_dag}{$v'$}
                \EndFor
            \EndIf
        \EndIf
        \EndUpon
    
        \vspace{0.4em}
        \Upon {receiving $\tuple{vt, r}_p$}
            \If {$vt.round=r$} 
            \State $votes[r]\gets votes[r]\cup \{vt\}$
            \State $sigVote[r] \gets sigVote[r]\cup \{vt.sig\}$
            \EndIf
            \If{$\exists v'\in DAG_i[r-1]: v'.source=L_{r-1} \land vt.strongEdge = L_{r-1}$}
            \State $sigCommit[r] \gets vt.sig$ 
            \EndIf
        \EndUpon
        \vspace{0.4em}
        
        \Upon{receiving $\VC_{r}^p$}
                \State $sigVote[r] \gets sigVote[r] \cup \{\VC_{r}^p.sig\}$
        \EndUpon
        

        
        
        \Upon{receiving $\CVC_{r}^p$}
            \State $sigCommit[r]\gets sigCommit[r]\cup\{\CVC_{r}^p.sig\}$
        \EndUpon

        \vspace{0.4em}
        \Upon {
        $(|sigVote[r]| + |DAG_i[r]| \geq n-f) \land ((\exists v' \in DAG_i[r] : v'.source = L_r)\ \lor\ \TC_r$ is received for $round=r$)}\label{line: advance round}
        \If{$|DAG_i[r]|< n-f$}
            \State $\VC_{r}^i \gets$ \Call{AggregateBLS}{$sigVote[r]$} 
            \State multicast $\VC_{r}^i$
        \EndIf
        \If{$P_i=L_{r+1} \land \nexists v' \in DAG_i[r] : v'.source = L_r$}
            \State \textbf{wait until} receiving $\TC_{r}$
            \For{$r'=r-1$ down to $1$}
            \If{$\exists v'' \in DAG_i[r'] : v''.source = L_r'$}
                \State \textbf{break}
            \Else 
                \State \textbf{wait until} receiving $\TC_{r'}$
            \EndIf
            \EndFor
        \EndIf
        
        \State advance\_round$(r+1)$
        \EndUpon
        \vspace{0.4em}

        \Upon {$timeout$} \label{line: timeout}
        \If {$\not \exists v' \in DAG_i[round] : v'.source = L_{round}$}
        \State multicast $\langle\Timeout, round\rangle_i$
        \EndIf

            \vspace{0.4em}
            
        \EndUpon
         \Upon{receiving $\TC_r$ such that $r\ge round$}
        \State multicast $\TC_r$
        \EndUpon

            \algstore{bkbreak}
    \end{algorithmic}
    \end{minipage}
    \hfill
   \begin{minipage}{0.48\textwidth}
    \begin{algorithmic}
    \algrestore{bkbreak}
    
    \Procedure {try\_add\_to\_dag}{$v$}
    \If {$\forall v' \in v.strongEdges \cup v.weakEdges \cup v.leaderEdges: v' \in \bigcup_{k \geq 1} DAG_i[k]$}
        \State $DAG_i[v.round] \gets DAG_i[v.round] \cup \{v\}$
        \State $\textit{buffer}\gets \textit{buffer}\ \backslash\ \{v\}$
        \State \Return $\True$
    \EndIf
    \State \Return $\False$
    \EndProcedure
        
    \vspace{0.4em}
    
    \Procedure{create\_new\_vertex}{$r$}
    \State $v.round \gets r$
    \State $v.source \gets \node{i}$
    \State $v.block \gets blocksToPropose$.dequeue()

    \If {$r>1$}
    \If{$\tuple{\Timeout,r-1}_i$ is sent}
    \State $v.strongEdges \gets v.strongEdges\ \backslash\ \{v':v'.source=L_{r-1}\}$
    \EndIf
    \EndIf
    \If{$\node{i} = L_r$ $\land$ $\not \exists v'\in DAG_i[r-1]:v'.source=L_{r-1}$}
    \For {$r'=r$ down to $1$}
    \If{$\not \exists v'' \in v.strongEdges \text{ s.t. } v''.source =L_{r'-1}$}
    \State wait for \(\TC_{r'-1}\)
    \State $v.tc\gets v.tc\cup\{\TC_{r'-1}\}$
    \Else 
    \State $v^* \gets$ \Call{get\_leader\_vertex}{$r'$}
    \State $v.leaderEdge \gets v^*$
    \State \textbf{break}
    \EndIf
    \EndFor
    \EndIf
    \State \Call{set\_weak\_edges}{$v, r$}
    \State \Return $v$
    \EndProcedure

    \vspace{0.4em}
    
    \Procedure{create\_new\_vote}{$r$}
    \State $vt.round \gets r$
        \State $vt.source \gets \node{i}$
        \If {$\tuple{\Timeout,r-1}_i$ is sent}
        \State $vt.strongEdge\gets \bot$
        \Else
        \If {$r>1\land \exists v' \in DAG_i[r-1] : v'.source = L_{r-1}$}
        \State $vt.strongEdge\gets v'$
        \EndIf
        \EndIf
        \State \Return $\tuple{vt, r}_i$
    \EndProcedure

    \vspace{0.4em}
                
    \Procedure {advance\_round}{$r$}
        \State $round\gets r$; $start\ timer$
        \If{$P_i=L_r \lor willProposeVertex[r]=\True$}
            \State broadcast\_vertex($round$)
        \Else
            \State multicast\_vote($round$)
        \EndIf
    \EndProcedure

    \vspace{0.4em}
     
    \Upon {init}
    \State advance\_round($1$)
    \EndUpon
    \algstore{bkbreak}
    \end{algorithmic}
    \end{minipage}
\end{boxedminipage}
\caption{\name: DAG construction protocol for party $\node{i}$.}
\label{fig:dag_construction}
\end{figure*}

%% file: jump_round_fig.tex
\begin{figure}[!ht]
	\footnotesize
	\begin{boxedminipage}[t]{\columnwidth}
	
            \begin{algorithmic}
            \algrestore{bkbreak}

        \Upon {$round<r-1\land |sigVote[r]| +|DAG_i[r]|\geq f+1\land(\exists v' \in DAG_i[r] : v'.source = L_r\ \lor\TC_r$ is received$)$} \label{line: jumping round}
        \If{$P_i=L_{r+1} \land \nexists v' \in DAG_i[r] : v'.source = L_r$}
            \State \textbf{wait until} receiving $\TC_{r}$
            \For{$r'=r-1$ down to $1$}
            \If{$\exists v'' \in DAG_i[r'] : v''.source = L_r'$}
                \State \textbf{break}
            \Else 
                \State \textbf{wait until} receiving $\TC_{r'}$
            \EndIf
            \EndFor
        \EndIf
        \State advance\_round$(r)$
        \EndUpon
        
     

    \algstore{bkbreak}
    \end{algorithmic}
\end{boxedminipage}
\caption{\name: Jumping rounds.}
\label{fig:jumping round}
\end{figure}

%% file: commit_rule_fig.tex
\begin{figure}[!ht]
    \footnotesize
	\begin{boxedminipage}[t]{\columnwidth}
		\textbf{Local variables:}
		\setlist{nolistsep}
		\begin{itemize}[noitemsep]
			\item[] $committedRound \gets 0$
		\end{itemize}
		\begin{algorithmic}
            \algrestore{bkbreak}
        \Event{receiving a set $\BigW$ of first messages for round $r+1$ vertices s.t. $\forall v'\in\BigW:\exists v\in v'.strongEdges$ s.t. $v.source=L_r$ $\land$ $|\BigW|+|sigCommit[r]|\ge n-f$}\label{line: commit1}
        \State \Call{try\_commit}{$r, \BigW, sigCommit[r]$}
        \EndEvent
        \vspace{0.4em}
        \Procedure{try\_commit}{$r, \BigW, \BigS$}
            \State $v \gets$ \Call{get\_leader\_vertex}{$r$}
            \If{$committedRound < r$}
                \State \Call{commit\_leader}{$v$}
                \If{$|\{v' \in \BigW\mid$ \Call{strong\_edge}{$v', v$} $\}|<n-f$}
                \State $\CVC_{r}^i \gets$ \Call{AggregateBLS}{$\BigS$}
                \State multicast $\CVC_{r}^i$
                \EndIf
            \EndIf
        \EndProcedure

			\vspace{0.4em}
			
			\Procedure{commit\_leader}{$v$}
			\State $leaderStack.$\Call{push}{$v$}
			\State $r \gets v.round - 1$
			\State $v' \gets v$
			\While{$r > committedRound$}
			\State $v_s \gets$ \Call{get\_leader\_vertex}{$r$}
			\If{\Call{leader\_path}{$v', v_s$}}
			\State $leaderStack.$\Call{push}{$v_s$} \label{step:indirect-commit}
			\State $v' \gets v_s$
			\EndIf
			\State $r \gets r -1$
			\EndWhile
			\State $committedRound \gets v.round$
			\State \Call{order\_vertices}{$ $}
			\EndProcedure
        \algstore{bkbreak}
		\end{algorithmic}

	\end{boxedminipage}
	\caption{\name: The commit rule for party $\node{i}$.}
	\label{fig:commit-rule}
\end{figure}

%% file: multi-leader.tex
\section{Multi-leader \name in Depth} \label{sec: multi-leader}
In \name, the latency to commit the leader vertex is shorter than that for non-leader vertices. To reduce the commit latency of multiple vertices, we extend \name to support multiple leaders per round. In the best-case scenario, when all these leaders are honest, the respective leader vertices can be committed with a latency of one RBC plus $1\delta$.

\mypara{Multiple leaders in a round.} In this protocol, multiple leaders are chosen within a round based on the round number. One of these leaders serves as the main leader, while the others are designated as secondary leaders. The vertex proposed by the main leader is referred to as the main leader vertex, and the vertices proposed by the secondary leaders are termed secondary leader vertices. We introduce a $\NoVote$ message, and a $\NoVote$ certificate, denoted $\NVC$, is formed by a quorum of $\NoVote$ messages. In Multi-leader \name, the main leader’s responsibility is to ensure that the main leader vertex has a leader path to all leader vertices from previous rounds and to collect a $\NoVote$ certificate for any missing leader vertices.

To determine the multiple leaders in a given round, we define a deterministic function, get\_multiple\_leaders($r$), which returns an ordered list of leaders for round $r$. The first leader in this list serves as the main leader, while the subsequent leaders are designated as secondary leaders. As in \name, the main leader for round $r$ is denoted by $L_r$. We use $\ML_r$ to denote the ordered list of leaders provided by get\_multiple\_leaders($r$). $\ML_r[x]$ denotes the $x$-th element in the list, and $\ML_r[1]$ is the main leader. Additionally, $\ML_r[: x]$ represents the first $x$ leaders, while $\ML_r[x + 1 :]$ denotes the leaders in the list excluding the first $x$ leaders.

\mypara{DAG construction protocol.} The basic data structures differ slightly from those of \name. A vertex may contain references to multiple leader vertices, and a main leader vertex may additionally contain a $\NoVote$ certificate $\NVC$ for missing leader vertices. A vote message contains references to the main leader and the delivered secondary leaders. Round advancement remains gated by the main leader vertex or its timeout certificate; missing secondary leader vertices do not prevent parties from advancing.

Recall that in \name, each party $P_i$ waits for the round-$r$ leader vertex until a timeout. If the leader vertex is not delivered before the timeout, $P_i$ sends a $\tuple{\Timeout, r}$ message. Upon receiving either the round-$r$ leader vertex or $\TC_r$ (along with $n-f$ round-$r$ vertices and votes), $P_i$ advances to round~$r + 1$.
In Multi-leader \name, $P_i$ sends $\tuple{\Timeout, r}$ only when it does not deliver the round-$r$ main leader vertex before the timeout; it does not send $\Timeout$ messages when the secondary leader vertices are not delivered. Each party $P_i$ waits for the round-$r$ main leader vertex or $\TC_r$ (along with $n-f$ round-$r$ vertices and votes) to advance to round $r+1$.
Waiting for a $\TC$ for secondary leaders could allow a single faulty leader to slow down the protocol.
Here, the votes refer to vote messages only. Upon receiving a vote message, each party broadcasts it to ensure that all other parties also receive it. However, the main leader vertex must still provide proof for any undelivered leader vertices, which motivates the introduction of $\NoVote$ messages.

 After entering round $r+1$, $P_i$ sends $\tuple{\NoVote, \ell, r}$ for every $\ell\in \ML_r[2:]$ from which it did not receive the round-$r$ leader vertex.
 Unlike Multi-leader Sailfish~\cite{shrestha2024sailfish}, $\NoVote$ messages are broadcast rather than sent only to the round $r+1$ leader. This choice is motivated by the structure of leader paths: in \name, each leader has a leader path to some previous round leader; in Multi-leader \name, the main leader has leader paths to the previous-round leaders (or collects $\NoVote$ certificates for any missing ones). Thus, $\NoVote$ messages should be received by all subsequent leaders.

\begin{figure}[ht]
    \centering
    \includegraphics[height=4.5cm]{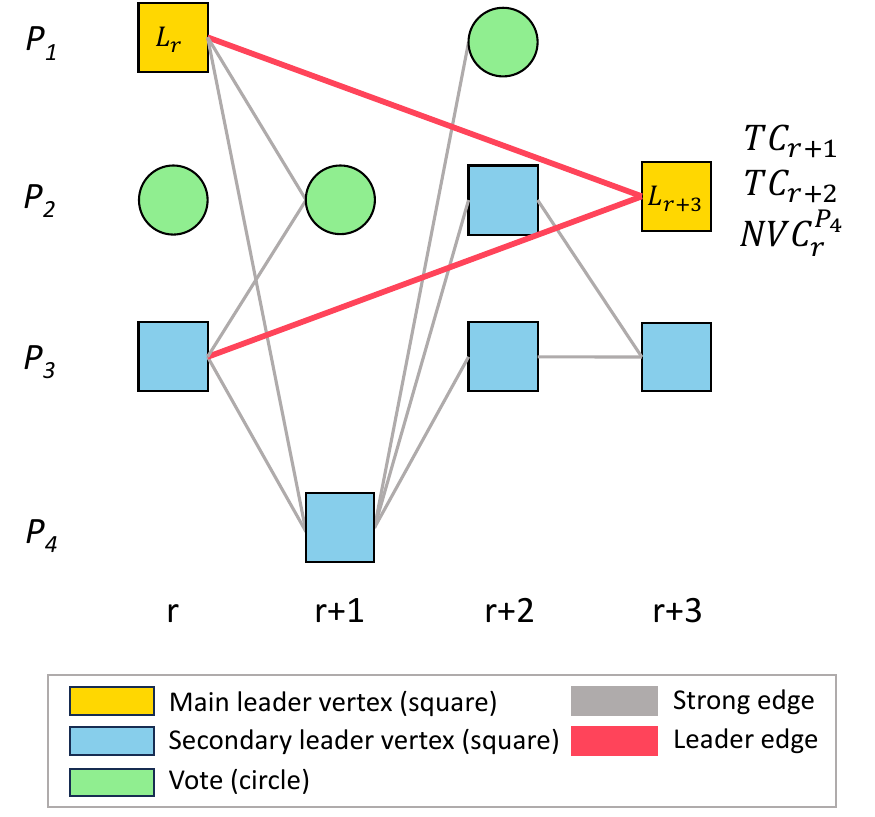}
    \caption{The set of local vertices and votes observed by $P_2$. In round $r$, the secondary leaders are $P_3$ and $P_4$. $L_{r+3}$ did not deliver $P_4$’s round-$r$ leader vertex or the round-$(r+1)$ and round-$(r+2)$ main leader vertices; therefore, $L_{r+3}$'s vertex must contain $\TC_{r+1}$, $\TC_{r+2}$, and $\NVC_{r}^{P_4}$.}
    \label{fig:multi-leader-illustration}
\end{figure}

We impose the following constraints on the main leader vertex. Suppose $L_{r'}$ is the last main leader that $L_{r+1}$ has delivered. (i) The round-$(r+1)$ main leader vertex must reference the round-$r'$ main leader vertex. (ii) If $L_{r+1}$ has not delivered the round-$r$ main leader vertex, then the round-$(r+1)$ main leader vertex $v_k$ must contain $\TC_{r''}$ for all rounds $r''$ such that $r'< r''<r$. (iii) The round-$(r+1)$ main leader vertex $v_k$ must establish leader paths to all leader vertices corresponding to leaders in $\ML_{r'}[:x]$, for some $x\ge 1$, and contain a $\NoVote$ certificate $\NVC_{r'}^{\ell}$ for leader $\ell=\ML_{r'}[x+1]$ and round $r'$. An illustration is provided in \Cref{fig:multi-leader-illustration}. If the main leader vertex has leader paths to all leader vertices corresponding to leaders in $\ML_{r'}$, it is not required to include $\NVC_{r'}$ for any round-$r'$ leaders. The is\_valid() function is also appropriately updated to ensure that these constraints are met. The constraint for other round-$(r+1)$ vertices remains unchanged. For round-$(r+1)$ vote messages, we require that they vote for all leaders and have strong edges to all delivered leader vertices if the round-$r$ main leader is delivered. If the round-$r$ main leader is not delivered, the strong edges to secondary leaders are not used in the commit rule and therefore are not required.

The complete pseudocode appears in the full paper~\cite{yu2025angelfish}.

\mypara{Committing and ordering the DAG.}
As in \name, only the leader vertices are committed, and the non-leader vertices are ordered (in some deterministic order) as part of the causal history of a leader vertex when the leader vertex is (directly or indirectly) committed, as implemented by the order\_vertices function in \Cref{fig:multi-leader commit rule}.
\input{multi_leader_commit_rule_fig}
In our protocol, an honest party $P_i$ directly commits a round-$r$ leader vertex $v_k$ corresponding to $\ML_r[x]$ when it observes a total of $n-f$ round-$(r+1)$ vertices and votes with strong edges to $v_k$ (counting the ``first messages'' of RBC for vertices) and when all round-$r$ leader vertices corresponding to leaders in $\ML_r[: x-1]$ have been directly committed. If $v_k$ fails to be directly committed, party $P_i$ refrains from committing the leader vertices corresponding to the leaders in $\ML_r[x+1 :]$, even if there are $n-f$ round-$(r+1)$ vertices and votes with strong paths to the leader vertices corresponding to the leaders in $\ML_r[x + 1 :]$. The commit rule is presented in the try\_commit() function (see \Cref{line: multi-leader try commit}).

Upon directly committing the main leader vertex $v_m$ in round $r$, $P_i$ first indirectly commits leader vertices corresponding to $\ML_{r'}[: y]$ (for some $y > 0$) in an earlier round $r' < r$ such that there exist leader paths from $v_m$ to all leader vertices corresponding to $\ML_{r'}[: y]$. Subsequently, $P_i$ repeats this process for leader vertices of earlier rounds that have strong paths from the leader vertex corresponding to $\ML_{r'}[1]$, until it reaches a round $r'' < r$ in which it previously directly committed a leader vertex (see \Cref{line: multi-leader commit leaders}). When round-$r'$ leader vertices corresponding to leaders in $\ML_{r'}[: y]$ are directly committed, we ensure that any future main leader vertex has a leader path to these round-$r'$ leader vertices. This ensures that these leader vertices will be (directly or indirectly) committed by honest parties that missed directly committing them.

%% file: multi_leader_commit_rule_fig.tex
\begin{figure}[!ht]
    \footnotesize
	\begin{boxedminipage}[t]{\columnwidth}
		\textbf{Local variables:}
		\setlist{nolistsep}
		\begin{itemize}[noitemsep]
			\item[] $committedRound \gets 0$
		\end{itemize}
		\begin{algorithmic}
            \algrestore{bkbreak}
     
        \Procedure{try\_commit}{$r, \BigS$} \label{line: multi-leader try commit}
            \State $\CLS \gets [\ ]$
            \For {$\ell \in \ML_r$} \label{line: multi-leader commit}
                \State $v\gets$ \Call{get\_vertex}{$\ell,r$}
                \If{$|collect| \ge n-f \land committedRound < r$}
                \State $\CLS\gets\CLS\ \Vert\ v$
                \Else\ \textbf{break}
            \EndIf
            \EndFor
            \State \Call{commit\_leaders}{$\CLS$}
        \EndProcedure

			\vspace{0.4em}
			
			\Procedure{commit\_leaders}{$cls$} \label{line: multi-leader commit leaders}
			\State $leaderStack.$\Call{push}{$cls$}
                \State $v'\gets cls[0]$
			\State $r \gets v'.round - 1$
			\While{$r > committedRound$}
                \State $\CMV \gets [\ ]$
                \For{$\ell\in\ML_r$}
                \State $v\gets$ \Call{get\_vertex}{$\ell,r$}
                \If{\Call{leader\_path}{$v', v$}}
                \State $\CMV\gets \CMV\ \Vert\ v$
                \Else\ \textbf{break}
			\EndIf
                \EndFor
                \If {$\CMV\neq [\ ]$}
                \State $v'\gets \CMV[1]$\Comment{main leader vertex for round $r$}
                \EndIf
			\State $leaderStack.$\Call{push}{$\CMV$} 
			\State $r \gets r -1$
			\EndWhile
			\State $committedRound \gets cls[0].round$
			\State \Call{order\_vertices}{$ $}
			\EndProcedure
            \vspace{0.4em}
    \algstore{bkbreak}  
    \end{algorithmic}
    
    \begin{algorithmic}[1]
    \algrestore{bkbreak}
    \Procedure{order\_vertices}{$ $}
    \While{$\neg leaderStack$.isEmpty()}
        \State $\CMV \gets leaderStack$.pop()
        \For{$v \in \CMV$} \Comment{iterate over $\CMV$ in order}
            \State $verticesToDeliver \gets \{v' \in \bigcup_{r>0} DAG_i[r] \mid path(v, v') \land v' \notin deliveredVertices\}$
            \For{\textbf{every} $v' \in verticesToDeliver$ in some deterministic order}
                \State \textbf{output} a\_deliver${_i}$($v'.block, v'.round, v'.source$)
                \State $deliveredVertices \gets deliveredVertices \cup \{v'\}$
            \EndFor
        \EndFor
    \EndWhile
\EndProcedure
\algstore{bkbreak}
    \end{algorithmic}
\end{boxedminipage}
\caption{Multi-leader \name: commit rule.}
\label{fig:multi-leader commit rule}
\end{figure}

%% file: evaluation.tex
\section{Empirical Evaluation}\label{sec: evaluation}

We implement \name and evaluate its performance on a real-world testbed by comparing it with the state-of-the-art DAG-based protocol Sailfish~\cite{shrestha2024sailfish} and Autobahn~\cite{giridharan2024autobahn}, a state-of-the-art protocol with asynchronous data dissemination.
We use Sailfish as the main DAG-based baseline because it achieves lower latency than earlier DAG-based protocols such as Tusk~\cite{danezis2022narwhal} (three-RBC latency) and Shoal~\cite{spiegelman2023shoal} (two-RBC latency), and we base \name's implementation on the Sailfish implementation;
this makes it much easier to perform controlled experiments with confidence that differences in performance are due to the algorithmic techniques deployed rather than idiosyncrasies of the codebases (e.g. networking code optimization, data structures used, etc.).
We do not compare against Mysticeti~\cite{babel2023mysticeti} because its implementation is based on a substantially different codebase, making a meaningful comparison difficult.
Unfortunately, no implementation of a state-of-the-art leader-based protocol that uses AVID (e.g., DispersedSimplex~\cite{shoup2024sing}) is available.
Instead, we obtain a baseline lower bound of 220 ms on the latency achievable by such protocols as follows: we use Hydrangea~\cite{shresthaHydrangeaOptimisticTwoRound2025}, a latency-optimal (\(3\delta\)), leader-based protocol, which we configure to use no mempool and empty blocks.
In addition, we evaluate the performance of Multi-leader \name as well as \name under failure scenarios.

Our implementation modifies the core consensus logic of the open-source Sailfish codebase~\cite{sailfish-impl} to create \name\footnote{https://github.com/qyu100/Angelfish/tree/Angelfish} and Multi-leader \name\footnote{https://github.com/qyu100/Angelfish/tree/Angelfish\_multi\_leader}. 

\mypara{Experimental setup.} We conducted our evaluations on the Google Cloud Platform (GCP), deploying nodes evenly across five distinct regions: us-east1-b (South Carolina), us-west1-a (Oregon), europe-west1-b (Belgium), europe-north1-b (Finland), and asia-northeast1-a (Japan). We employed e2-standard-16 instances~\cite{gcp}, each featuring $16$ vCPUs, $64$ GB of memory, and up to 16 Gbps of network bandwidth. All nodes ran on Ubuntu $22.04$, and round-trip latencies between GCP regions ranged from roughly 30 ms to 250 ms (details appear in~\Cref{tab:ping-latencies}). Sailfish and \name use BLS signatures.

\begin{table}[ht]
\centering
\caption{Ping latencies (in ms) between GCP regions.}
\small 
\label{tab:ping-latencies}
\begin{threeparttable}
\begin{tabular}{l|c r r r r}
\toprule
\multicolumn{1}{c|}{} & \multicolumn{5}{c}{\textbf{Destination}*} \\
\midrule
\textbf{Source} & us-e1 & us-w1 & eu-w1 & eu-n1 & as-n1 \\
\midrule
us-east1-b              & 0.70 & 63.95 & 93.05 & 115.44 & 157.61 \\
us-west1-a              & 65.03 & 0.69 & 135.15 & 157.31 & 89.59 \\
europe-west1-b          & 93.06 & 135.18 & 0.55 & 31.25 & 223.37 \\
europe-north1-b         & 115.21 & 156.29 & 31.22 & 0.77 & 248.25 \\
asia-northeast1-a       & 167.57 & 89.60 & 228.51 & 249.32 & 0.67\\
\bottomrule
\end{tabular}
\begin{tablenotes}
\small
\item[*] Region names are abbreviated versions of the source regions.
\end{tablenotes}
\end{threeparttable}
\end{table}

In our evaluations, each party generates a configurable number of transactions ($512$ random bytes each) for inclusion in its vertex, with a vertex containing up to $10,000$ transactions (i.e., $5$ MB). Each experiment runs for $120$ seconds with $50$ nodes or $60$ seconds with $10$ nodes. The timeout parameter $\tau$ is set to $5$ seconds. Latency is measured as the average time between the creation of a transaction and its commit by all non-faulty nodes. Throughput is measured by the number of committed transactions per second.

\begin{figure}[ht]
    \centering
    \includegraphics[width=0.32\textwidth]{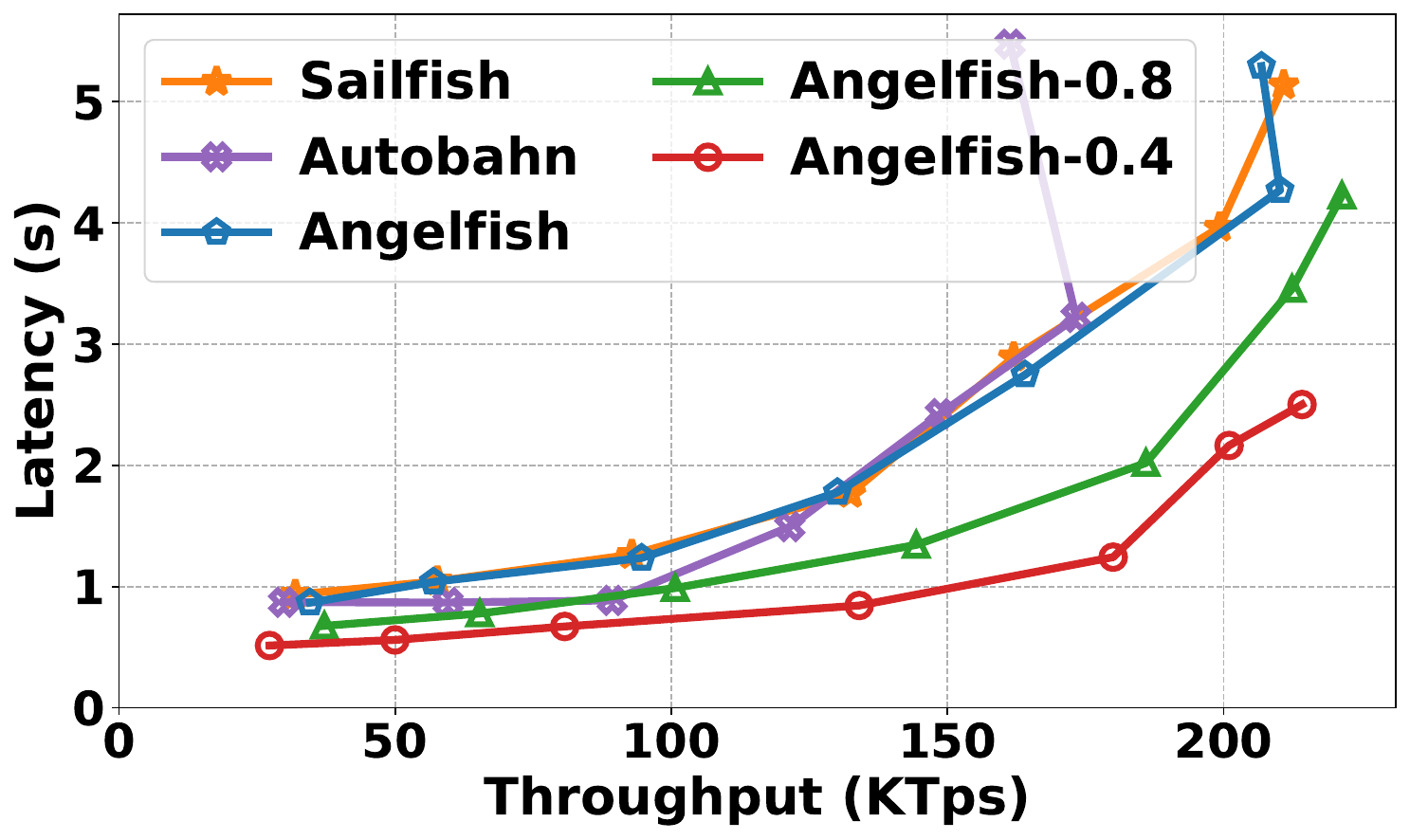}
    \caption{Throughput vs. latency with $50$ nodes.}
    \label{fig:overall}
\end{figure}

\mypara{Methodology.} In our evaluations, we gradually increased the number of transactions per vertex. As shown in~\Cref{fig:overall}, throughput increases, accompanied by a slight increase in latency, up to an inflection point that quickly leads to saturation.

    
    
    

\begin{figure*}[htbp]
  \centering
  \begin{minipage}{0.32\textwidth}
    \centering
    \includegraphics[width=\linewidth]{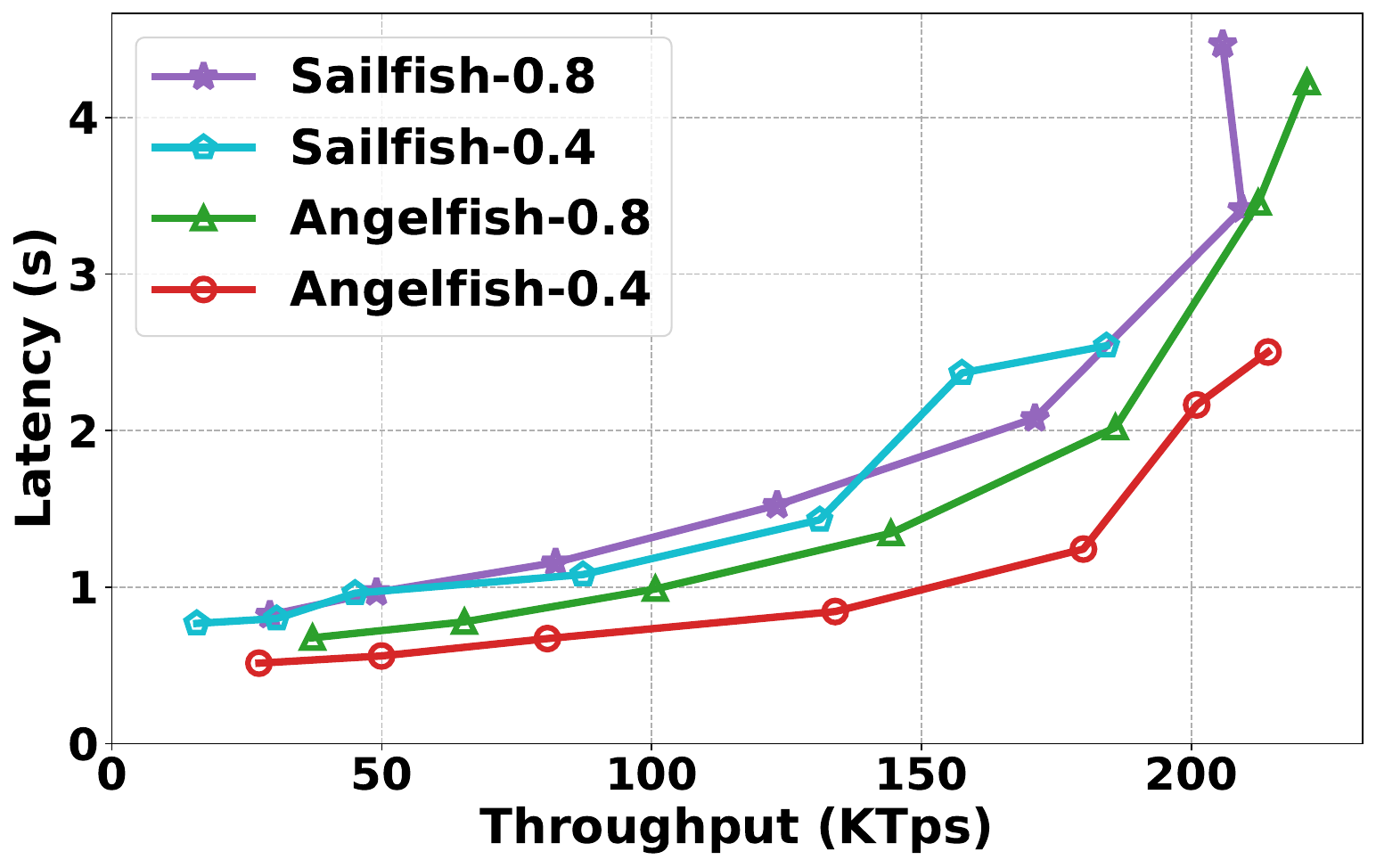}
    \caption{Using empty blocks in Sailfish vs.\ \name. Throughput vs. latency at various propose\_rate values with $50$ nodes.}
    \label{fig:ideal}
  \end{minipage}
  \hfill
  \begin{minipage}{0.32\textwidth}
    \centering
    \includegraphics[width=\linewidth]{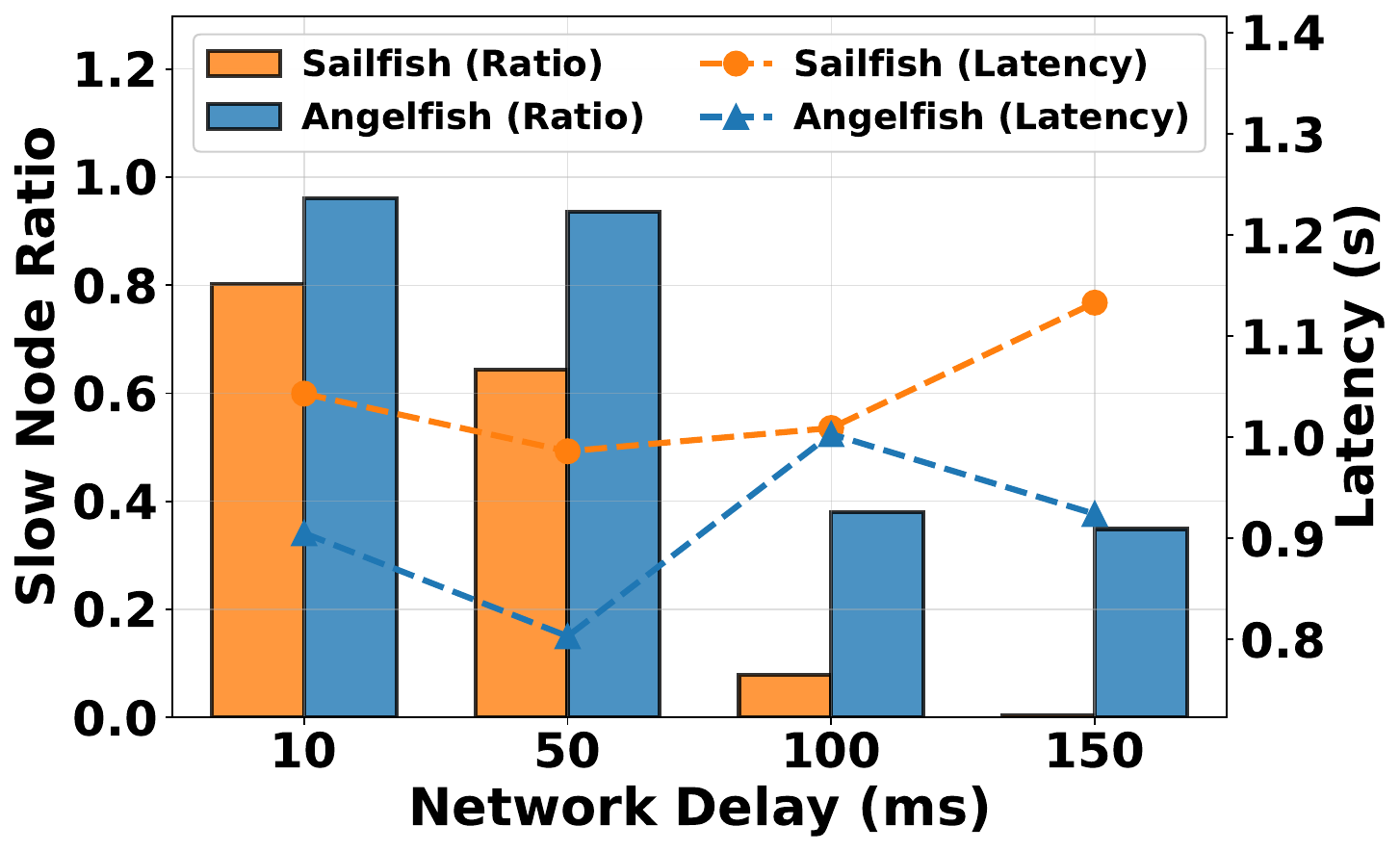}
    \caption{Slow node ratio vs. network delay with $50$ nodes.}
    \label{fig:slow}
  \end{minipage}
  \hfill
  \begin{minipage}{0.32\textwidth}
    \centering
    \includegraphics[width=\linewidth]{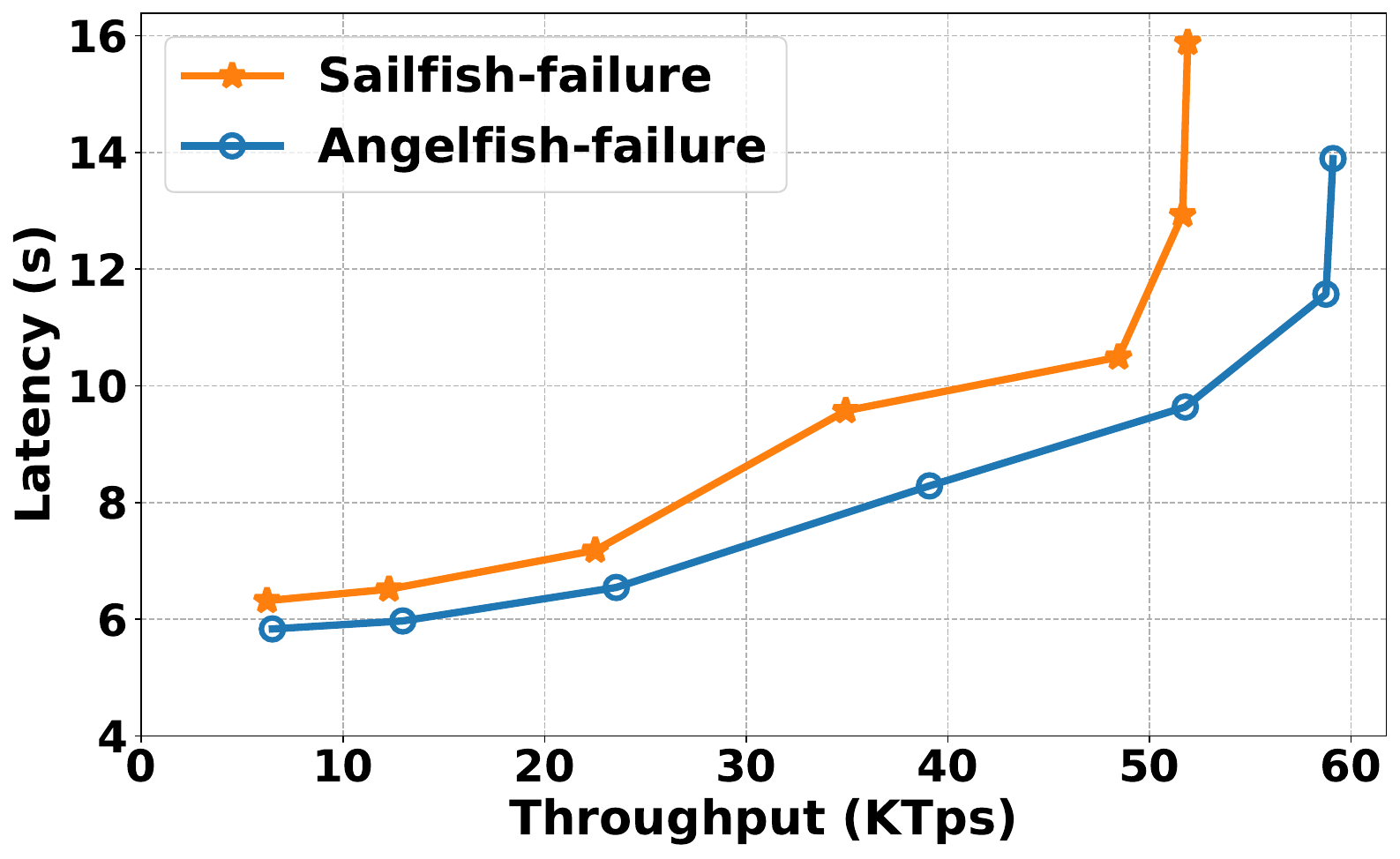}
    \caption{Throughput vs. latency at $50$ nodes with 16 crash failures.}
    \label{fig:failure}
  \end{minipage}
\end{figure*}

\mypara{Performance comparison in fault-free cases.}
We compare the performance of Sailfish, Autobahn, and \name under fault-free scenarios. In \name, the parameter propose\_rate denotes the fraction of nodes that propose vertices in each round, while the other nodes only vote. \Cref{fig:overall} presents the latency of Sailfish and \name as throughput increases from left to right, with \name evaluated under propose\_rate values of $1$, $0.8$, and $0.4$.

In the implementation, we fix the propose\_rate for each round and randomly select vertex proposers according to this rate. If the leader is not among the selected proposers, we additionally require the leader to propose a vertex. As a result, each round has either $50 \times$ propose\_rate or $50 \times$ propose\_rate $+ 1$ vertex proposers.
Both \name and Sailfish employ two-round RBC and adopt an optimistic signature verification strategy: nodes aggregate signatures without verifying each one individually and verify only the final aggregate.

When propose\_rate equals $1$, \name and Sailfish perform nearly indistinguishably; this is expected since the two protocols behave identically in this scenario. When the propose\_rate is $0.8$ or $0.4$, \name achieves lower latency than Sailfish, since a higher propose\_rate requires nodes to wait for proposals from more regions and reduces the number of early one-round votes, both of which increase latency. When throughput is lower than 120\,KTps, Autobahn performs better than Sailfish. However, its peak throughput is lower than those of Sailfish and Angelfish.


In the case of propose\_rate$=0.8$, $40(41)$ nodes propose vertices in each round. In \name, a party must deliver at least the number of proposers it observes minus the fault tolerance, i.e., $40(41) - 16 = 24(25)$, in order to enter the next round.
As a result, each party waits for between $24$ and $40$ vertices, corresponding to proposals from the nearest $3$--$5$ regions. If the threshold of $n-f$ is not reached by the time the leader vertex is delivered, it gathers vote messages to advance to the next round. In the case of $\text{propose\_rate}=0.4$, the minimum number of vertices a node must wait for is $50 \times 0.4 - 16 = 4$, which means nodes only need to wait for proposals from the nearest $1$ to $5$ regions. Additionally, more nodes use vote messages than with $\text{propose\_rate}=0.8$. Thus, the latency of $\text{propose\_rate}=0.4$ is lower than that of $\text{propose\_rate}=0.8$.


Finally, with a $\text{propose\_rate}$ of $0.4$, \name processes roughly 27\,KTps at 0.5\,s latency, which is only about twice the lower-bound latency we measure when Hydrangea orders empty blocks.

\mypara{Comparing BEB of vote messages to RBC of empty vertices.}
An important question is whether \name{}'s BEB votes have an advantage compared to just proposing empty vertices in Sailfish.
To find out, we set up Sailfish\footnote{https://github.com/qyu100/Angelfish/tree/Sailfish\_add\_propose\_rate} so that only a fraction of the nodes, given by \(\text{propose\_rate}\), propose vertices containing transaction blocks while the others propose empty vertices. 
We then compared Angelfish and Sailfish for \(\text{propose\_rate}\) values of \(0.4\) and \(0.8\).

The results appear in \Cref{fig:ideal}.
We can see that both protocols achieve roughly the same peak throughput regardless of the propose rate.
However, \name achieves consistently lower latency (at least 100 ms) at all throughput levels.
This is consistent with our claim that BEB votes arrive faster than RBC vertices, even when empty, allowing nodes to reach the quorum threshold and move to the next round faster.

\mypara{Tolerance to slower nodes.} 
\Cref{fig:slow} illustrates that Angelfish exhibits greater tolerance of slow nodes than Sailfish. In our experiment, we introduced artificial network delays (implemented via the \texttt{tc} command) to 16 nodes uniformly distributed across 5 regions, designating them as slow nodes. In Angelfish\footnote{https://github.com/qyu100/Angelfish/tree/Angelfish\_slow\_nodes}, these nodes only contribute votes, whereas in Sailfish\footnote{https://github.com/qyu100/Angelfish/tree/Sailfish\_slow\_nodes}, they propose empty vertices. Leaders in both protocols are selected in a round-robin manner from the remaining 34 normal nodes. Each block contains a fixed batch of 1,000 transactions (i.e., 0.5~MB). Each experiment runs for 120 seconds. We then measured slow-node participation per consensus round. Specifically, we counted how many of the first $2f+1$ votes (in Angelfish) or vertices (in Sailfish) received came from the slow nodes.
Results appear in~\Cref{fig:slow}, where the $x$-axis represents the artificial network delay and the $y$-axis represents the percentage of votes or vertices from slow nodes that are useful, i.e., received among the first \(2f+1\) messages supporting a leader.
The results show that Angelfish allows more slow nodes to meaningfully participate than Sailfish. This is because votes are sent in a single communication step in Angelfish, whereas vertices require two steps in Sailfish.


\mypara{Performance comparison under failures.}
\Cref{fig:failure} presents the performance of \name\footnote{https://github.com/qyu100/Angelfish/tree/Angelfish\_failure} and Sailfish\footnote{https://github.com/qyu100/Angelfish/tree/Sailfish\_failure} with 16 crash failures among $50$ nodes.
We designate a failed leader every three rounds, and the $16$ crashed nodes are uniformly distributed across five regions. The propose\_rate is set to $1$.


In Sailfish, after the timer expires, each party receives a $\Timeout$ certificate, enters a new round, and sends a $\NoVote$ message to the leader. The new leader must then collect a quorum of $\NoVote$ messages before proposing a vertex. In contrast, \name requires only the $\Timeout$ certificate to proceed. This optimization eliminates one message delay compared to Sailfish, which explains the latency gap observed in the figure. As the crash frequency increases, this gap will become more pronounced.

\begin{figure}[ht]
    \centering
    \includegraphics[width=0.32\textwidth]{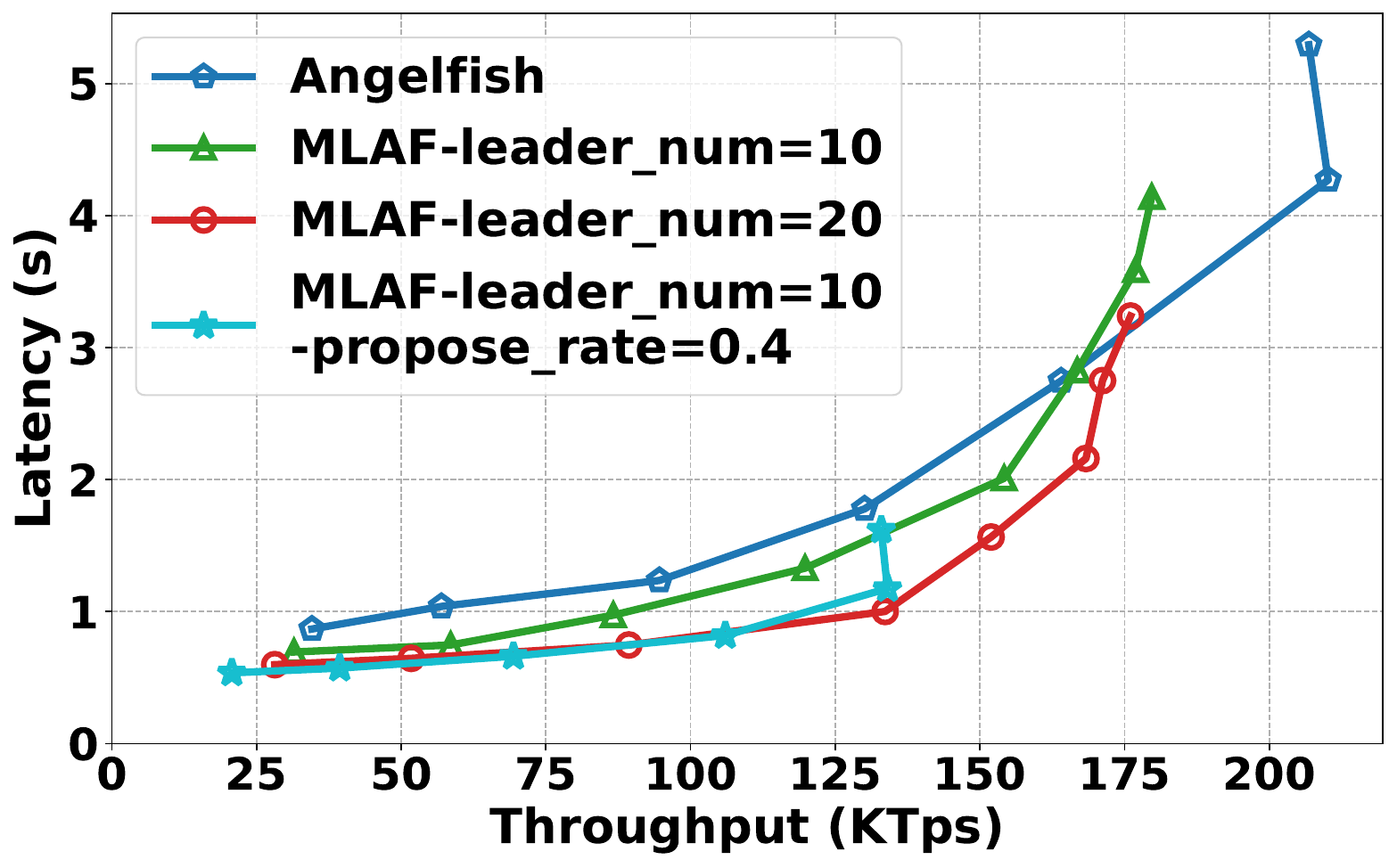}
    \caption{Latency comparison of Multi-leader \name in the absence of failures with $50$ nodes.}
    \label{fig:multi}
\end{figure}

\mypara{Performance comparison of Multi-leader \name.}
We evaluated Multi-leader \name in failure-free settings with $10$ and $20$ leaders per round. The results are presented in \Cref{fig:multi}. In the figure, MLAF-leader\_num=$x$ denotes configurations with $x$ leaders, while MLAF-leader\_num=$10$-propose\_rate=$0.4$ indicates the case with $10$ leaders in which $60\%$ of parties abstain from proposing vertices.

As shown in the figure, consensus latency improves as the number of leader vertices increases.
Moreover, with propose\_rate=$0.4$, Multi-leader \name is now only 36\% slower than the lower bound obtained with Hydrangea proposing empty blocks.

\mypara{Performance comparison of the small-timeout design.}
We evaluate two variants\footnote{https://github.com/qyu100/Angelfish/tree/Angelfish\_small\_timeout} that use a small fixed timeout of $25$~ms or $50$~ms. The fixed timeout increases the average number of parents from $4.5$ to $5.27$ under the $25$~ms timeout and to $5.5$ under the $50$~ms timeout (the analysis was conducted with a 0.5 MB input, represented by the second point of each line), allowing more non-leader vertices to be committed within two RBC delays. However, this benefit does not offset the extra waiting time on the commit path: the two timeout rounds add roughly $50$~ms and $100$~ms of latency, and the net latency still increases by about $40$~ms and $70$~ms, respectively.
\begin{figure}[ht]
    \centering
    \includegraphics[width=0.32\textwidth]{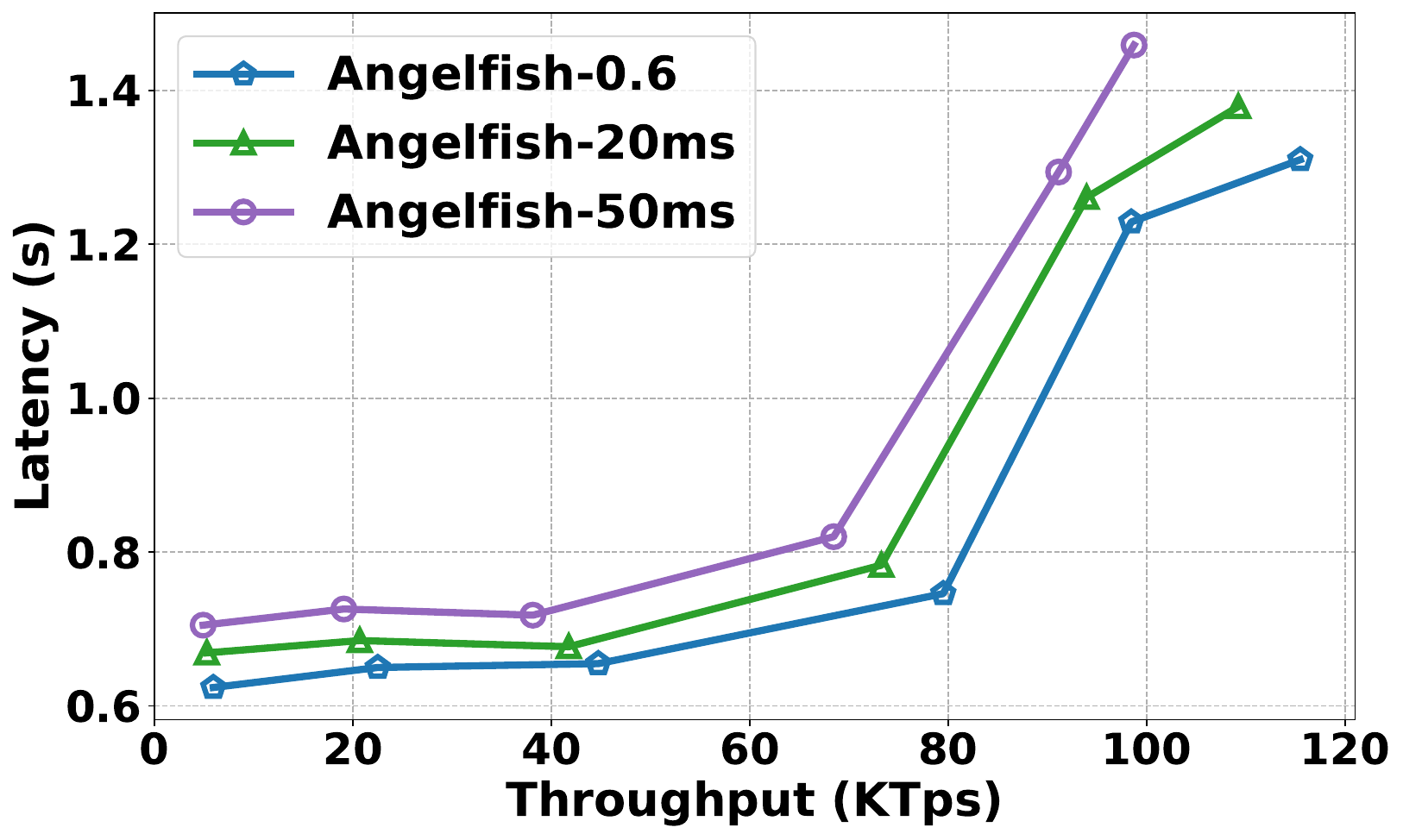}
    \caption{Performance comparison of small timeouts with $10$ nodes.}
    \label{fig:small_timeout}
\end{figure}

%% file: related_work.tex
\section{Related Work}\label{sec: related work}

In the vein of PBFT~\cite{castro1999practical}, traditional leader-based protocols~\cite{gueta2019sbft,kotla2007zyzzyva,martinFastByzantineConsensus2006a,suri-payerBasilBreakingBFT2021,yu2024tetrabft} use a rotating leader that is responsible for broadcasting data and driving consensus.
This achieves low latency, but throughput is limited by the leader's bandwidth.
Simplex~\cite{chanSimplexConsensusSimple2023} exemplifies this design in its simplest form: in each round, the leader broadcasts a block and the other parties vote on it.
To achieve higher throughput, newer protocols try to make use of the bandwidth of all parties to disseminate data.

Some protocols like RCC~\cite{guptaRCCResilientConcurrent2021}, MIR-BFT~\cite{stathakopoulou2019mir}, or ISS~\cite{stathakopoulouStateMachineReplication2022} spread the load by operating parallel protocol instances that each use a different leader and combine their outputs.
SpotLess~\cite{kang2024spotless} follows this concurrent rotational line of work and combines independent chained consensus instances with rapid view synchronization to avoid single-leader bottlenecks and simplify recovery.
These protocols parallelize leader-based consensus instances, whereas \name keeps the DAG-based dissemination structure and reduces the latency of DAG formation by allowing parties without transactions to send lightweight votes instead of reliably broadcasting empty vertices.

Protocols like Destiny~\cite{arunScalableByzantineFault2022}, Jolteon~\cite{gelashvili2022jolteon}, or Autobahn~\cite{giridharan2024autobahn} reliably store data using an asynchronous data-dissemination layer and separately perform consensus on data references.
This achieves high throughput but adds data-storage latency to consensus latency.
Autobahn is able to linearize entire causal histories at once after network blips.

Another line uses asynchronous verifiable information dispersal (AVID) based on erasure codes to distribute the dissemination load~\cite{cachin2005asynchronous,yang2022dispersedledger,shoup2024sing}.
Using AVID, DispersedSimplex~\cite{shoup2024sing}, a variant of Simplex, retains the latency-optimized structure of partially synchronous leader-based protocols while addressing the leader bottleneck, although erasure coding introduces processing, communication, and data-expansion costs.

In DAG-based protocols, parties create data blocks in parallel and organize them in a DAG whose structure is then interpreted to arrive at a total ordering.
HashGraph~\cite{baird2016swirlds} pioneered this approach with an unstructured, asynchronous DAG.
Aleph~\cite{gkagol2019aleph} first introduced a layered DAG built in rounds. DAG-Rider~\cite{keidarAllYouNeed2021a} further optimizes performance and resilience and is quantum-safe. Narwhal~\cite{danezis2022narwhal} showed that the approach achieves high performance in practice.
Bullshark~\cite{spiegelman2022bullshark} introduces leader vertices and timeouts in the DAG-building process in exchange for better latency in the common case, and it is the first partially synchronous DAG-based protocol with embedded consensus.
Shoal~\cite{spiegelman2023shoal} runs multiple instances of Bullshark in a pipeline to further reduce latency.
Shoal++~\cite{arun2025shoal++} makes further improvements upon Bullshark by using multiple leader vertices per round and intertwining the outputs of parallel protocol instances.
Sailfish~\cite{shrestha2024sailfish} is the first to achieve a leader vertex every round and commit leader vertices in 3 message delays.
Sailfish++~\cite{shrestha2025optimistic} is the first DAG-based BFT protocol to use signature-free reliable broadcast.

Uncertified DAG-based protocols such as Cordial Miners~\cite{keidar2023cordial}, BBCA-Chain~\cite{malkhi2023bbca}, Mysticeti~\cite{babel2023mysticeti}, and Black Marlin~\cite{ignacio2025dag} use best-effort broadcast to disseminate DAG vertices instead of reliable broadcast.
This avoids incurring the \(O(n^2)\) complexity of reliable broadcast (per vertex) but may result in brittle performance~\cite{arun2025shoal++}, as missing vertices must be fetched later on the critical consensus path.

Fully asynchronous BFT protocols avoid timing assumptions. A classic line follows the Ben-Or--Kelmer--Rabin asynchronous common-subset framework~\cite{benor1994asynchronous}, combining reliable broadcast with asynchronous binary agreement to select proposals. HoneyBadger and BEAT~\cite{miller2016honey,duan2018beat} follow or optimize this construction.

While the protocols above achieve dramatically higher throughput than traditional leader-based protocols, the latter still reign over latency.
In concurrent work, Morpheus~\cite{lewis2025morpheus} and Grassroots Consensus~\cite{keidar2025grassroots} attempt to solve this problem by dynamically switching between a low-throughput mode that behaves like a traditional leader-based protocol and a high-throughput mode that uses decoupled data dissemination.
In contrast, \name is able to adapt continuously along this spectrum to achieve the best tradeoff at any point.

%% file: conclusion.tex
\section{Conclusion and Future Work}\label{sec:conclusion}

\name can be seen as a hybrid between Simplex~\cite{chanSimplexConsensusSimple2023}, a leader-based protocol, and Sailfish~\cite{shrestha2024sailfish}, a DAG-based protocol.
When only round leaders propose vertices and all other parties only vote, \name behaves similarly to Simplex, except that it piggybacks the next leader proposal onto the leader's vote.
When all nodes propose vertices, \name behaves similarly to Sailfish, except that, in asynchronous periods, leaders' edges may point several rounds back.
In both cases, unlike Simplex, \name can take advantage of all nodes that have transactions to propose to increase throughput and, unlike Sailfish, nodes that do not have transactions to propose incur less latency overhead thanks to \name's lightweight votes.
Lightweight votes may transfer to other DAG-based protocols, but each protocol will need its own safety argument: \name's depends on the precise commit rule, timeout mechanism, and DAG-validity conditions.
We leave a systematic exploration of this design space for future work.

%% file: Open_Science.tex
\section*{Open Science and Ethical Considerations}

Our implementations of \name, of Multi-leader \name, and of the modified Sailfish baselines are at \url{https://github.com/qyu100/Angelfish}, with the branch of each experiment given in the footnotes of \Cref{sec: evaluation}; the TLA+ specification and its model-checking configuration are archived at \url{https://doi.org/10.5281/zenodo.21855535}.
This work involves no human subjects, no personal or third-party data, and no experiments on systems we do not operate, and we see no risk of harm arising from its publication or from the release of our artifacts.

%% file: RBC_denotion.tex
\section{Extended Preliminaries} \label{sec: RBC denotion}
\begin{definition}[Byzantine reliable broadcast~\cite{bracha1987asynchronous}]
In a Byzantine reliable broadcast (RBC), a designated sender $\node{k}$ may invoke \Call{r\_bcast$_k$}{$m$} to propagate an input $m$. Each party $\node{i}$ may then output (we also say commit) the message $m$ via \Call{r\_deliver$_i$}{$m, \node{k}$} where $\node{k}$ is the designated sender. The reliable broadcast primitive satisfies the following properties:
\begin{itemize}[noitemsep,leftmargin=*]
    \item[-] \textbf{Validity.} If the designated sender $\node{k}$ is honest and calls \Call{r\_bcast$_k$}{$m$}, then every honest party eventually outputs \Call{r\_deliver}{$m,\node{k}$}.
    \item[-] \textbf{Agreement.} If an honest party $\node{i}$ outputs \Call{r\_deliver$_i$}{$m, \node{k}$}, then every other honest party $\node{j}$ eventually outputs \Call{r\_deliver$_j$}{$m, \node{k}$}.
    \item[-] \textbf{Integrity.} Each honest party $\node{i}$ outputs \Call{r\_deliver$_i$}{} at most once regardless of $m$.
\end{itemize}
\end{definition}

%% file: security_analysis.tex
\section{Security Analysis}\label{sec: angelfish security analysis}
We say that a \textit{leader vertex $v_i$ is committed directly} by party $\node{i}$ if $\node{i}$ invokes \Call{commit\_leader}{$v_i$}. Similarly, we say that a \textit{leader vertex $v_j$ is committed indirectly} if it is added to $leaderStack$ in Line~\ref{step:indirect-commit}. In addition, we say party $\node{i}$ consecutively directly commits leader vertices $v_k$ and $v_{k'}$ if $\node{i}$ directly commits $v_k$ and $v_{k'}$ in rounds $r$ and $r'$ respectively and does not directly commit any leader vertex between $r$ and $r'$.

The following fact is immediate from the use of reliable broadcast to propagate a vertex $v$, along with the requirement to wait for its causal history of vertices to be added to the DAG before adding $v$.
\begin{fact}\label{fact}
For every two honest parties \( P_i \) and \( P_j \) (i) for every round \( r \), \( \bigcup_{r' \leq r} DAG_i[r'] \) is eventually equal to \( \bigcup_{r' \leq r} DAG_j[r'] \), (ii) At any given time \( t \) and round \( r \), if \( v \in DAG_i[r] \) and  \( v' \in DAG_j[r] \) such that \( v.source = v'.source \), then \( v = v' \). Moreover, for every round \( r' < r \), if \( v'' \in DAG_i[r'] \) and there is a path from \( v \) to \( v'' \), then \( v'' \in DAG_j[r'] \) and there is a path from \( v' \) to \( v'' \).
\end{fact}

\begin{claim} \label{claim: leader path}
    If an honest party \( P_i \) directly commits a round-$r$ leader vertex $v_k$, then for every round \( r' \) leader vertex \( v_\ell \)  such that \( r' > r \), there exists a leader path from \( v_\ell \) to $v_k$.
\end{claim}
\begin{proof}
    If an honest party directly commits a round-$r$ leader vertex $v_k$, it must have observed $n-f$ ``support'' signals in total, including both (i) first RBC messages from round-$(r{+}1)$ vertices that reference the round-$r$ leader vertex, and (ii) votes (vote messages $\tuple{vt, r+1}_*$ that support it and signatures from $\CVC_{r+1}$). By quorum intersection, there cannot be $n-f$ vertices and votes that do not have a strong edge to $v_k$. If an honest party $P_j$ has sent a $\tuple{\Timeout, r}_j$, it will not propose a round-$(r+1)$ vertex or vote that has a strong edge to it, so $\TC_r$ will not form. Thus, a leader edge from a higher-round $>r$ leader vertex to a lower-round $<r$ leader vertex will not exist. In our protocol, every leader vertex must have a leader edge or strong edge referencing a leader vertex from some previous round, thereby forming a continuous leader path. Since $v_k$ cannot be skipped, any leader vertex $v_\ell$ must have a leader path to $v_k$.
\end{proof}

\begin{claim}\label{claim: commit}
    If an honest party \( P_i \) directly commits a leader vertex $v_k$ in round \( r \) and an honest party $P_j$ directly commits a leader vertex \( v_\ell \) in round \( r'\ge r \), then $P_j$ (directly or indirectly) commits $v_k$ in round \( r \).
\end{claim}

\begin{proof}
If $r' = r$, by Fact~\ref{fact}, $v_k = v_\ell$ and $P_j$ directly commits $v_k$. When $r' > r$, by Claim~\ref{claim: leader path}, $v_\ell$ has a leader path to $v_k$. By the code of commit\_leader, after directly committing a leader vertex in round $r'$,  $P_j$ tries to commit a leader vertex if there exists a leader path between the two leader vertices from a smaller round until it meets the last round $r'' < r'$ that directly committed a leader vertex. If $r'' < r < r'$, $P_j$ must indirectly commit $v_k$ in round $r$. If $r < r''$, by inductive argument and~\Cref{claim: leader path}, $P_j$ must indirectly commit $v_k$ after directly committing round-$r''$ leader vertex.
\end{proof}

\begin{claim} \label{claim: same order}
    Let $v_k$ and \( v'_k \) be two leader vertices consecutively directly committed by a party
     \( P_i \) in rounds \( r_i \) and \( r'_i > r_i \) respectively.
     Let \( v_\ell \) and \( v'_\ell \) be two leader vertices consecutively directly committed by
     party \( P_j \) in rounds \( r_j \) and \( r'_j > r_j \) respectively.
     Then, \( P_i \) and \( P_j \) commit the same leader vertices between rounds \( \max(r_i, r_j) \)
     and \( \min(r'_i, r'_j) \), and in the same order.
\end{claim}

\begin{proof}
If $r'_i < r_j$ or $r'_j < r_i$, then there are no rounds between $\max(r_i, r_j)$ and $\min(r'_i, r'_j)$ and we are trivially done. Otherwise, assume without loss of generality that \( r_i \leq r_j < r_i' \leq r_j' \). By Claim~\ref{claim: commit}, both \( P_i \) and \( P_j \) will (directly or indirectly) commit the same leader vertices in round \( \min(r_i', r_j') \). Assume without loss of generality that \( \min(r'_i, r'_j) = r'_i \). By Fact~\ref{fact}, both \( {DAG}_i \) and \( {DAG}_j \) will contain \( v'_k \) and all leader vertices have a path from \( v'_k \) in \( {DAG}_i \). According to the commit\_leader procedure, after directly or indirectly committing the leader vertex \( v'_k \), each party will attempt to indirectly commit leader vertices from earlier rounds by leader path until reaching a round in which they have previously directly committed a leader vertex. Consequently, both \( P_i \) and \( P_j \) will indirectly commit the leader vertices in the leader path from rounds \( \min(r'_i, r'_j) \) down to \( \max(r_i, r_j) \). Furthermore, due to the fixed logic of the commit\_leader code, both parties will commit the same leader vertices between rounds \( \min(r'_i, r'_j) \) and \( \max(r_i, r_j) \) in the same order.
\end{proof}

By inductively applying~\Cref{claim: same order} for every pair of honest parties, we get the following:
\begin{corollary} \label{corollary: same order}
    Honest parties commit the same leader vertices in the same order.
\end{corollary}

\begin{lemma}[Total order] \label{lemma: total order}
    If an honest party $P_i$ outputs $a\_deliver_i$ $(b, r, P_k)$ before
    $a\_deliver_i(b', r', P_\ell)$, then no honest party $P_j$ outputs
    $a\_deliver_j(b', r', P_\ell)$ before $a\_deliver_j(b, r, P_k)$.
\end{lemma}

\begin{proof}
    By~\Cref{corollary: same order}, all honest parties commit the same leader vertices in the same order. According to the logic of order\_vertices, parties process committed leader vertices in that order and a\_deliver all vertices in their causal history based on a predefined rule. By~\Cref{fact}, every honest party has an identical causal history in their DAG for each committed leader. This establishes the lemma.
\end{proof}

\begin{lemma}[Agreement] \label{lemma: agreement}
    If an honest party $\node{i}$ outputs $a\_deliver_i$ $(v_i.block, v_i.round,$ $v_i.source)$, then every other honest party $P_j$ eventually outputs $a\_deliver_j$ $(v_i.block, v_i.round,$ $v_i.source)$.
\end{lemma}

\begin{proof}
   Since $\node{i}$ outputs a\_deliver$_i(v_i.block, v_i.round,$ $v_i.source)$, it follows from the order\_vertices logic that there exists a leader vertex $v_k$ committed by $\node{i}$ such that $v_i$ is in the causal history of some leader vertex $v_k$. When $P_i$ eventually commits leader vertex $v_k$, by \Cref{lemma: total order}, $P_j$ also commits $v_k$. By \Cref{fact}, the causal histories of $v_k$ in $DAG_i$ and $DAG_j$ are the same. Thus, when $\node{j}$ orders the causal histories of $v_k$, it outputs $a\_deliver_j(v_i.block, v_i.round,$ $v_i.source)$.
\end{proof}

\begin{lemma}[Integrity] \label{lemma: integrity}
    For every round $r\in \mathbb{N}$ and party $P_k \in \mathcal{P}$,
    an honest party $P_i$ outputs $a\_deliver_i(b,r,P_k)$ at most once regardless of $b$.
\end{lemma}

\begin{proof}
    An honest party $P_i$ outputs a\_deliver$_i(v.block, v.round, v.$ $source)$ only when vertex $v$ is in $DAG_i$. Note that $v$ is added to $P_i$'s DAG upon the reliable broadcast r\_deliver$_i(v.block, v.round, v.$ $source)$ event. Therefore, the proof follows from the Integrity property of reliable broadcast.
\end{proof}

\mypara{Validity.} We rely on GST to prove validity. Additionally, we use RBC protocol of Bracha~\cite{bracha1987asynchronous} to establish validity. Bracha's RBC protocol operates in $3$ communication steps and ensures the RBC properties at all times. After GST, it offers the following stronger guarantees:

\begin{property} \label{property: 3Delta}
    Let $t$ be a time after GST. If an honest party reliably broadcasts a message $M$ at time $t$, then all honest parties will deliver $M$ by time $t+3\Delta$.
\end{property}

\begin{property} \label{property: 2Delta}
    Let $t_g$ denote the GST. If an honest party delivers message $M$ at time $t$, then all honest parties deliver $M$ by time $\max(t_g, t)+2\Delta$.
\end{property}



\begin{claim} \label{claim: optimistic}
    Let $t_g$ denote the GST and $\node{i}$ be the first honest party to enter round $r+1$. If $\node{i}$ enters round $r+1$ at time $t$, then all honest parties will enter round $r$ or higher by $\max(t_g, t) + 2\Delta$.
\end{claim}
\begin{proof}
Observe that by time $t$, $\node{i}$ must have delivered round-$(r-1)$ vertices and received $r-1$ votes (vote messages and vote certificates), with their total count being $n-f$. Since when an honest party enters a round by votes, it will send vote certificates, then all honest parties must have received the vote certificates by $\max(t_g, t) + \Delta$. By \Cref{property: 2Delta}, all honest parties must have delivered round-$(r-1)$ vertices by $\max(t_g, t) + 2\Delta$.

We then prove the claim by considering $\node{i}$ entering round by delivering round-$r$ leader vertex (say $v_k$) and by receiving $\TC_{r-1}$. If $P_i$ delivered $v_k$, by~\Cref{property: 2Delta}, all honest parties must have delivered $v_k$ by $\max(t_g, t) + 2\Delta$. Otherwise, $P_i$ must have multicasted $\TC_{r-1}$ which arrives at all honest parties by $\max(t_g, t) + \Delta$. For $L_r$, if it enters round $r$, it also has to deliver round-$r'$ leader vertex and receive $\TC$s of rounds between $r'$ and $r$. Similarly, by $\max(t_g, t) + 2\Delta$ it will deliver round-$r'$ leader vertex and by $\max(t_g, t) + \Delta$ it will receive these $\TC$s, since for each round, either leader vertex is delivered, $\TC$ exists or both. Thus, all honest parties will enter round $r$ by $\max(t_g, t) + 2\Delta$ if they have not already entered a higher round.
\end{proof}

\begin{claim} \label{claim: keep entering}
    All honest parties keep entering higher rounds.
\end{claim}
\begin{proof}
    Suppose all honest parties are in round $r$ or higher. Let party $\node{i}$ be in round \( r \). If there exists an honest party \( P_j \) in a round \( r' > r \) at any time, then by Claim~\ref{claim: optimistic}, all honest parties will eventually enter round \( r' \) or higher. Otherwise, all honest parties remain in round \( r \). Note that upon entering round \( r \), all honest parties first compute the number of parties that will propose a vertex in round $r-1$, say $c$, based on the round-$(r-2)$ vertices and votes. Then they wait for $\max(c-f,0)$ round-$(r-1)$ vertices. After that, all honest parties r\_bcast a round \( r \) vertex, or broadcast vote certificates so each party will deliver round-$r$ vertices and receive votes with a total number of $n-f$.

    If no honest party has delivered the round \( r \) leader vertex,
    then by the timeout rule, all honest parties will multicast \( \langle \Timeout, r \rangle \) and receive $\TC_r$. In addition, if $L_{r+1}$ delivers a round $r'< r$ leader vertex, it will also receive $\TC_{r'+1},...\TC_{r}$, so that there is a leader edge to round-$r'$ leader vertex. Thus, all honest parties will move to round $r+1$. On the other hand, if at least one honest party has delivered the round \( r \) leader vertex, then by Fact~\ref{fact}, all honest parties will eventually deliver this round \( r \) leader vertex as well. With round \( r \) vertices and the round \( r \) leader vertex delivered and votes received with a total number of vertices and votes being $n-f$, all honest parties will proceed to round \( r+1 \).
\end{proof}

\begin{claim} \label{claim: honest entry}
    If an honest party enters round $r$ and creates a vertex, then at least $f + 1$ honest parties must have already entered round~$r-1$.
\end{claim}

\begin{proof}
    If an honest party $P_i$ enters round $r$ and creates a vertex, it must do so either directly from round $r-1$ or by jumping rounds from some round $< r-1$. In either case, there must be at least one honest party that has delivered round-$(r - 1)$ vertices and received $\tuple{vt,r-1}_*$ or vote certificates with a total number of $n-f$. At least $f+1$ of these vertices and votes are sent by honest parties while they are in round $r-1$. Thus, $f+1$ honest parties must have already entered~$r - 1$.
\end{proof}

\begin{claim}\label{claim: GST strong path}
    If the first honest party to enter round $r$ does so after GST and $L_r$ is honest and proposes, then every honest party creates a round-$(r+1)$ vertex or votes with a strong edge to the round-$r$ leader vertex.
\end{claim}
\begin{proof}
    We prove the claim by considering the case where all honest parties set their timeout parameter $\tau = 5\Delta$. Let $t$ be the time when the first honest party (say $\node{i}$) entered round $r$. Observe that no honest party sends $\sig{\Timeout, r}$ before $t+5\Delta$ due to its round timer expiring. Thus, $\TC_r$ does not exist before $t+5\Delta$. Additionally, by Claim~\ref{claim: honest entry}, no honest party can enter a round greater than $r$ until at least $f + 1$ honest parties have entered \( r \). Thus, no honest party sends a $\Timeout$ message for a round greater than \( r \) before $t + 5\Delta$ and no honest party enters a higher round by receiving a $\Timeout$ certificate before $t + 5\Delta$.

    Since $\node{i}$ entered round $r$ at time \( t \), by Claim~\ref{claim: optimistic}, all honest parties will enter round \( r \) or higher by \( t + 2\Delta \). If any honest party enters a round higher than $r+1$ before \( t + 5\Delta \), there exist at least $n-f$ round-$(r+1)$ vertices and votes. This is because, for an honest party to enter round $r'$, there must exist at least one honest party that has delivered round-$(r'-1)$ vertices and received $\tuple{vt, r'-1}_*$ with a total number of $n-f$. By transitive argument, there must exist $n-f$ round-$(r+1)$ vertices and votes. Moreover, no honest party sends a $\tuple{\Timeout, r}$ message before $t+ 5\Delta$. Since an honest party does not send a round-$(r+1)$ vertex or $\tuple{vt,r+1}_*$ without a strong edge to round-$r$ leader vertex (say $v_k$) without receiving $\TC_r$, this implies $n-f$ round-$(r+1)$ vertices and votes must have a strong edge to $v_k$.

    Also, note that if an honest party enters round $r + 1$ before $t+5\Delta$, there must be an honest party that has delivered $n-f$ round-$r$ vertices and received $\tuple{vt,r}_*$ with a total number of $n-f$, along with vertex $v_k$ (since $\TC_r$ does not exist before $t+5\Delta$). Thus, its round-$(r + 1)$ vertex or vote must have a strong edge to $v_k$.

    Now consider the case where no honest party has entered a round higher than $r$ before time $t + 5\Delta$. By Claim~\ref{claim: optimistic}, all honest parties enter round $r$ by $t + 2\Delta$. Upon entering round $r$, each honest party invokes r\_bcast on its round-$r$ vertex or multicasts its vote. By Property~\ref{property: 3Delta}, round-$r$ vertices from all honest parties are delivered by $t + 5\Delta$. Thus, all honest parties in round $r$ will deliver round-$r$ vertices and receive $\tuple{vt, r}_*$ with a total number being $n-f$ along with $v_k$ and send round-$(r+1)$ vertices or votes with a strong edge to $v_k$.
\end{proof}

\begin{claim}\label{claim:can-propose}
    If the first honest party to enter round $r$ does so after GST and both $L_r$ and $L_{r+1}$ are honest, then \(L_{r+1}\) proposes in round~\(r+1\).
\end{claim}
\begin{proof}
    We must show that \(L_{r+1}\) is able to create a leader edge to some round \(r' \leq r\).
    The only problematic case is if \(L_{r+1}\) sent a timeout message in round \(r\) but it does not have a timeout certificate for round \(r\).
    In that case, the rules of the protocol do not allow it to propose: it cannot create a leader edge to round \(r\) because it sent a timeout message in round \(r\), and it also cannot skip over round \(r\) because it does not have a timeout certificate for round \(r\).

    However, this cannot happen if the leader of round \(r\) is honest and round \(r\) started after GST.
    There are two cases.
    First, suppose that \(L_r\) proposed in round \(r\).
    Then, by~\Cref{claim: GST strong path}, \(L_{r+1}\) does not time out in round \(r\) and the problem does not arise.
    Second, suppose that \(L_r\) does not propose in round \(r\).
    Then a timeout certificate will form for round \(r\) and \(L_{r+1}\) will receive it and the problem will not arise.
\end{proof}

By the commit rule and Claims \ref{claim: GST strong path} and \ref{claim:can-propose}, we get the following:

\begin{corollary}\label{corollary: validity}
    If the first honest party to enter round $r$ does so after GST and $L_r$ and $L_{r+1}$ are honest, then all honest parties will directly commit the leader vertex of round \(r+1\).
\end{corollary}

\begin{lemma}[Validity]~\label{lemma: validity}
    If an honest party $\node{i}$ calls $a\_bcast_i(b, r)$, then every honest party eventually outputs $a\_deliver_i(b, r, P_i)$.
\end{lemma}

\begin{proof}
    When an honest party \( P_i \) calls a\_bcast$_i(b, r)$, it pushes \( b \) into the \( blocksToPropose \) queue.
    We now show that every block in this queue is eventually committed and delivered.
    By Claim~\ref{claim: keep entering}, \( P_i \) continuously progresses to higher rounds, creating new vertices in each of these rounds.
    Because leaders are assigned pseudorandomly to rounds, there will come a round \(r\) such that \(r-1\) started after GST, the leader of round \(r-1\) is honest, and \(P_i\) is the leader of round \(r\).
    In round \(r\), \( P_i \) will dequeue an element \(b\) from \( blocksToPropose \) and create a vertex \( v_i \) with block \( b \), and by~\Cref{corollary: validity}, all honest parties will commit it and output $a\_deliver_i(b, r, P_i)$.
\end{proof}

%% file: multi-leader_security_analysis.tex
\section{Multi-leader \name Security Analysis}\label{sec: multi-leader appendix}

We say that a leader vertex $v_i$ is committed directly by party $P_i$ if $P_i$ invokes commit leaders($\CLS$) and $v_i \in \CLS$. Similarly, we say that a leader vertex $v_j$ is committed indirectly if $\CMV$ is added to $leaderStack$ and $v_j \in \CMV$. In addition, we say party $P_i$ consecutively directly commits leader vertices in rounds $r$ and $r' > r$ and does not directly commit any leader vertex between $r$ and $r'$.

\begin{claim} \label{claim: multi-leader leader path}
If an honest party $P_i$ directly commits a leader vertex $v_k$ in round $r$, then for every main leader vertex $v_\ell$ in round $r'$ such that $r' > r$, there exists a leader path from $v_\ell$ to $v_k$.
\end{claim}

\begin{proof}
If $v_k$ is a main leader vertex, the proof is the same as \Cref{claim: leader path}. We now consider the case where $v_k$ is not a main leader vertex.

If an honest party directly commits a round-$r$ leader vertex $v_k$, it must have observed $n-f$ messages in total, including both (i) first RBC messages from round-$(r+1)$ vertices that reference the round-$r$ leader vertex, and (ii) vote messages $\tuple{vt, r+1}_*$ that support it.
By quorum intersection, it is impossible for $n-f$ such vertices and votes to lack a strong edge to $v_k$; equivalently, $n-f$ $\NoVote$ messages for $v_k$ cannot exist. By the commit rule, if $v_k$ is committed, then all leaders in the round-$r$ leader list that precede $v_k$ are also committed (see \Cref{line: multi-leader commit}). Similarly, $\NoVote$ messages for these vertices $\mathcal{H}$ do not exist. Thus, a leader edge from a higher-round $>r$ main leader vertex to a lower-round $<r$ leader vertex will not exist. In our protocol, every main leader vertex must have a leader edge or a strong edge referencing a main leader vertex from some previous round, thereby forming a continuous leader path. Since $\mathcal{H}$ cannot be skipped, any main leader vertex $v_\ell$ must have a leader path to $v_k$.
\end{proof}

The indirect commit rule of a main leader vertex in Multi-leader \name is identical to the indirect commit rule of the leader vertex in \name. Thus, the proof of the following claim (\Cref{claim: multi-leader main leader commit}) remains identical to \Cref{claim: commit} except \Cref{claim: multi-leader leader path} needs to be invoked (instead of \Cref{claim: leader path}).

\begin{claim} \label{claim: multi-leader main leader commit}
If an honest party $P_i$ directly commits the main leader vertex $v_k$ in round $r$ and an honest party $P_j$ directly commits the main leader vertex $v_\ell$ in round $r' \geq r$, then $P_j$ (directly or indirectly) commits $v_k$ in round $r$.
\end{claim}

\begin{claim} \label{claim: multi-leader commit}
If an honest party $P_i$ directly commits all leader vertices corresponding to $\ML_r[: x]$ (for some $x > 0$) and an honest party $P_j$ directly commits the main leader vertex $v_\ell$ in round $r' > r$, then $P_j$ indirectly commits all leader vertices corresponding to $\ML_r[: x]$ in round $r$.
\end{claim}

\begin{proof}
    Given that $P_i$ directly committed all leader vertices in $\ML_r[: x]$, by \Cref{fact} and \Cref{claim: multi-leader main leader commit}, there are leader paths from the main leader vertex of any round higher than $r$ to all leader vertices corresponding to $\ML_r[: x]$ in $DAG_j$. 
    
    By the code of commit\_leaders(), after directly committing the main leader vertex $v_\ell$ in round $r'$, $P_j$ tries to indirectly commit all leader vertices corresponding to $\ML_{r''}[: y]$ (for some $y > 0$) in an earlier round $r'' < r'$ such that there exist leader paths from $v_\ell$ to all leader vertices corresponding to $\ML_{r''}[: y]$.
    This process of indirectly committing multiple leader vertices of an earlier round is repeated for leader vertices that have leader paths from the main leader vertex of round $r''$ (i.e., $\ML_{r''}[1]$), until it reaches a round $r^* < r'$ in which it previously directly committed a leader vertex. If $r^* < r < r'$, party $P_j$ will indirectly commit all leader vertices in $\ML_r[: x]$ in round $r$. Otherwise, by inductive argument and \Cref{claim: multi-leader leader path}, party $P_j$ must have indirectly committed all leader vertices in $\ML_r[: x]$ when directly committing the main leader vertex of round $r^*$.
\end{proof}

\begin{claim} \label{claim: multi-leader same order}
Let an honest party $P_i$ consecutively directly commit in rounds $r_i$ and $r'_i$. Also, let an honest party $P_j$ consecutively directly commit in rounds $r_j$ and $r'_j$. Then, $P_i$ and $P_j$ commit the same leader vertices between rounds \( \max(r_i, r_j) \)  and $\min(r'_i, r'_j)$ and in the same order.
\end{claim}

\begin{proof}
If $r'_i < r_j$ or $r'_j < r_i$, then there are no rounds between $\max(r_i, r_j)$ and $\min(r'_i, r'_j)$ and we are trivially done. Otherwise, assume wlog that $r_i \le r_j < r'_i$. Also, assume $\min(r'_i, r'_j) = r'_i$. Let $\ML_{r'_i}[: x]$ be the list of multiple leader vertices directly committed by party $P_i$ in round $r'_i$ for some $x > 0$. If $r'_i = r'_j$, by Claim~\ref{claim: multi-leader main leader commit}, party $P_j$ commits at least $\ML_{r'_i}[1]$ in round $r'_i$. Otherwise, by Claim~\ref{claim: multi-leader commit}, party $P_j$ indirectly commits all leader vertices in $\ML_r[: x]$ in round $r'_i$.

Moreover, by Fact~\ref{fact}, both $DAG_i$ and $DAG_j$ will contain $\ML_{r'_i}[1]$ (i.e., the main leader vertex in round $r'_i$) and all vertices that have a path from $\ML_{r'_i}[1]$ in $DAG_i$. By the code of commit\_leaders(), after (directly or indirectly) committing $\ML_{r'_i}[1]$, parties try to indirectly commit multiple leader vertices in a smaller round number $r'' < r'_i$ that have leader paths from $\ML_{r'_i}[1]$. This process is repeated by indirectly committing leader vertices of earlier rounds with leader paths from $\ML_{r'_i}[1]$ until it reaches a round $r^* < r$ in which it previously directly committed a leader vertex. Therefore, both $P_i$ and $P_j$ will indirectly commit all leader vertices from $\min(r'_i, r'_j)$ to $\max(r_i, r_j)$. Moreover, due to the deterministic code of commit\_leaders, both parties will commit the same leader vertices between rounds $\min(r'_i, r'_j)$ and $\max(r_i, r_j)$ in the same order.
\end{proof}

By inductively applying \Cref{claim: multi-leader same order} between any two pairs of honest parties we obtain the following corollary.

\begin{corollary} \label{corollary: multi-leader same order}
Honest parties commit the same leaders in the same order.
\end{corollary}

The proof of the following total order lemma (\Cref{lemma: multi-leader total order}) remains identical to \Cref{lemma: total order} except \Cref{corollary: multi-leader same order} needs to be invoked (instead of \Cref{corollary: same order}).

\begin{lemma}[Total order] \label{lemma: multi-leader total order}
Multi-leader \name satisfies Total order.
\end{lemma}

\textbf{Agreement.} The agreement proof remains identical to \Cref{lemma: agreement} except \Cref{lemma: multi-leader total order} needs to be invoked (instead of \Cref{lemma: total order}).

\textbf{Integrity.} The integrity proof remains identical to \Cref{lemma: integrity}.

\textbf{Validity.} We again rely on GST to prove validity and utilize the RBC protocol from Bracha RBC~\cite{bracha1987asynchronous}.

\begin{claim} \label{claim: multi-leader enter rounds}
 Let $t_g$ denote the GST and $P_i$ be the first honest party to enter round $r$. If $P_i$ enters round $r$ at time $t$, then (i) all honest parties (except $L_r$ when $P_i \neq L_r$) enter round $r$ or higher by $\max(t_g, t)+2\Delta$, and (ii) $L_r$ (if honest and $P_i \neq L_r$) enters round $r$ or higher by $\max(t_g, t) + 4\Delta$.
\end{claim}

\begin{proof} 
Observe that by time $t$, $\node{i}$ must have delivered round-$(r-1)$ vertices and received $r-1$ votes, with their total count being $n-f$. Since when an honest party receives a vote, it will broadcast the vote, then all honest parties must have received these votes by $\max(t_g, t) + \Delta$. By \Cref{property: 2Delta}, all honest parties must have delivered round-$(r-1)$ vertices by $\max(t_g, t) + 2\Delta$. We then prove part (i) of the claim by considering $\node{i}$ entering round by delivering round-$r$ main leader vertex (say $v_k$) and by receiving $\TC_{r-1}$. If $P_i$ delivered $v_k$, by~\Cref{property: 2Delta}, all honest parties must have delivered $v_k$ by $\max(t_g, t) + 2\Delta$. Otherwise, $P_i$ must have multicasted $\TC_{r-1}$ which arrives at all honest parties by $\max(t_g, t) + \Delta$.

Having delivered $v_k$ or received $\TC_{r-1}$ (along with $n-f$ round-$(r-1)$ vertices and votes), an honest party $P_j$ broadcasts $\tuple{\NoVote, P_k, r-1}$ for all $P_k \in \ML_{r-1}$ if $P_j$ did not deliver its corresponding leader vertex by then. If no honest party delivered the leader vertex corresponding to $P_k$ by $\max(t_g, t) + 2\Delta$, then all honest parties (including $L_r$) will broadcast $\tuple{\NoVote, P_k, r-1}$. Thus, $L_r$ will receive $\NVC^{P_k}_{r-1}$ by time $\max(t_g, t) + 3\Delta$. For $L_r$, if it does not deliver round-$(r-1)$ main leader vertex, it must instead deliver the main leader vertex of some earlier round $r' < r$, together with the $\TC$s of all rounds between $r'$ and $r$, as well as the corresponding $\NVC$. By \Cref{property: 2Delta}, $L_r$ delivers the round-$r’$ vertices by $\max(t_g, t) + 2\Delta$. It receives $\TC_{r’}$ and the corresponding $\NVC$ by $\max(t_g, t) + \Delta$. On the other hand, if at least one honest party delivered the leader vertex corresponding to $P_k$ by $\max(t_g, t) + 2\Delta$, by \Cref{property: 2Delta}, $L_r$ will deliver the leader vertex corresponding to $P_k$ by $\max(t_g, t) + 4\Delta$. Thus, $L_r$ will either deliver a leader vertex corresponding to $P_k$ or receive $\NVC^{P_k}_{r-1}$ for all $P_k \in \ML_{r-1}$ by time $\max(t_g, t) + 4\Delta$. Since $L_r$ waits for leader vertices corresponding to $\ML_{r'}[:x]$ and $\NVC^p_{r'}$ where $r'<r$ and $p = \ML_{r'}[x+1]$, $L_r$ enters round $r$ by $\max(t_g, t) + 4\Delta$ if it has not already entered a higher round. This proves part (ii) of the claim.
\end{proof}

\begin{claim}
All honest parties keep entering increasing rounds.
\end{claim}

\begin{proof}
Suppose all honest parties are in round $r$ or above. Let party $P_i$ be in round $r$. If there exists an honest party $P_j$ in round $r' > r$ at any time, then by \Cref{claim: multi-leader enter rounds}, all honest parties will enter round $r'$ or higher. Otherwise, all honest parties are in round $r$. Upon entering round $r$, every honest party has observed messages from at least $n-f$ distinct round-$(r-1)$ participants, counting delivered vertices and received votes. It then r\_bcasts a round-$r$ vertex if it is a leader or has transactions to propose; otherwise, it multicasts a round-$r$ vote. Consequently, every honest party eventually delivers round-$r$ vertices and receives round-$r$ votes from at least $n-f$ distinct participants. Furthermore, if an honest party (except $L_{r+1}$) delivers the round-$r$ main leader vertex (say $v_k$), it will advance to round $r + 1$.

Alternatively, if no honest party delivered $v_k$ by the time their round-$r$ timer expires, due to the timeout rule, all honest parties will multicast $\langle \Timeout, r \rangle$ and subsequently receive $\TC_r$. Having delivered $v_k$ or received $\TC_r$, an honest party $P_j$ sends $\langle \NoVote, P_k, r \rangle$ for all $P_k \in \ML_r$ if $P_j$ did not deliver its corresponding leader vertex by then. If no honest party delivered the leader vertex corresponding to $P_k$ by the time they delivered $v_k$ or received $\TC_r$, then all honest parties will multicast $\langle \NoVote, P_k, r \rangle$. In addition, if $L_{r+1}$ delivers a round $r'< r$ leader vertex but does not deliver leader vertices from round $r'+1$ to round $r$, it will also receive $\TC_{r'+1},...\TC_{r}$, so that there is a leader edge to round-$r'$ leader vertex. Moreover, since $\langle \NoVote, P_k, r' \rangle$ for each $P_k \in \ML_{r'}$ is multicast, $L_{r+1}$ will also receive $\NVC^{P_k}_{r'}$. If all the leaders in round $r'$ are delivered, $L_{r+1}$ will also receive $\NVC^{P_k}_{r'+1}$ since $\langle \NoVote, P_k, r'+1 \rangle$ for each $P_k \in \ML_{r'+1}$ is multicast. By \Cref{property: 2Delta}, if at least one honest party delivered the leader vertex corresponding to $P_k$,  $L_{r+1}$ will deliver the leader vertex corresponding to $P_k$. Thus, one of the following cases occurs:
\begin{itemize}
    \item If $L_{r+1}$ delivers round-$r$ main leader vertex, then $L_{r+1}$ either delivers a leader vertex corresponding to $P_k$ in round $r$ or receives $\NVC^{P_k}_r$ for all $P_k \in \ML_r$. Since $L_{r+1}$ waits for leader vertices corresponding to $\ML_r[:x]$ and $\NVC^p_r$ where $p = \ML_r[x+1]$, it will advance to round $r+1$.
    \item If $L_{r+1}$ does not deliver round-$r$ main leader vertex, then it will deliver a leader vertex corresponding to $P_k$ in round $r'$, together with $\TC_{r'+1},...\TC_{r}$.
    \begin{itemize}
        \item If not all the leaders in round $r'$ are delivered, then $L_{r+1}$ receives $\NVC^{P_k}_{r'}$ for all $P_k \in \ML_{r'}$. Since $L_{r+1}$ waits for leader vertices corresponding to $\ML_{r'}[:x]$ and $\NVC^p_{r'}$ where $p = \ML_{r'}[x+1]$, $L_{r+1}$ will advance to round $r+1$. 
        \item If all the leaders in round $r'$ are delivered, then $L_{r+1}$ receives $\NVC^{P_k}_{r'+1}$ for all $P_k \in \ML_{r'+1}$. Since $L_{r+1}$ waits for $\NVC^p_{r'+1}$ where $p = \ML_{r'+1}[x+1]$, $L_{r+1}$ will advance to round $r+1$. 
    \end{itemize}
\end{itemize}
\end{proof}

The proof of the following claim (\Cref{claim: mult-leader GST strong path}) remains identical to \Cref{claim: GST strong path} except \Cref{claim: multi-leader enter rounds} needs to be invoked (instead of \Cref{claim: optimistic}).

\begin{claim} \label{claim: mult-leader GST strong path}
If the first honest party to enter round $r$ does so after GST and $L_r$ is honest, then there exists at least $n-f$ round-$(r + 1)$ vertices and votes with strong edges to round-$r$ main leader vertex.
\end{claim}

By the commit rule and \Cref{claim: mult-leader GST strong path}, the following corollary follows.

\begin{corollary} \label{corollary: multi-leader validity}
If the first honest party to enter round $r$ does so after GST and $L_r$ is honest, then all honest parties will directly commit the round-$r$ main leader vertex.
\end{corollary}

The proof of the following validity lemma (\Cref{lemma: multi-leader validity}) remains identical to \Cref{lemma: validity} except \Cref{corollary: multi-leader validity} needs to be invoked (instead of \Cref{corollary: validity}).

\begin{lemma}[Validity] \label{lemma: multi-leader validity}
Multi-leader \name satisfies Validity.
\end{lemma}

As demonstrated in \Cref{claim: mult-leader GST strong path}, a round-$r$ main leader vertex (proposed by an honest leader) is always committed by round $r + 1$ (after GST).

%% file: multi_leader_helper.tex
\section{Multi-leader \name Helper Function}\label{sec:multi-helper}
\input{multi_leader_dag_construction_fig}
\clearpage

%% file: multi_leader_dag_construction_fig.tex
\begin{figure*}[!ht]
	\footnotesize
	\begin{boxedminipage}[t]{\textwidth}
		\textbf{Local variables:}
		\setlist{nolistsep}
		\begin{itemize}[noitemsep]
			\item[] struct vertex $v$: \Comment{The struct of a vertex in the DAG}
			      \begin{itemize}[noitemsep]
			      	\item[] $v.round$ - the round of $v$ in the DAG
			      	\item[] $v.source$ - the party that broadcasts $v$
			      	\item[] $v.block$ - a block of transactions   
                        \item[] $v.strongEdges$ - a set of vertices in $v.round-1$ that represent strong edges
    				\item[] $v.weakEdges$ - a set of vertices in rounds $<$ $v.round-1$ that represent weak edges
                        {\color{blue}
                        \item[] $v.leaderEdges$ - leader vertices in rounds $<$ $v.round-1$ that represent leader edges (for leader vertices)
                        }
                        \item[] $v.tc$ - a set of $\Timeout$ certificates for round $\le$ $v.round-1$ (if any)
                        {\color{blue}
                        \item[] $v.nvc$ - a $\NoVote$ certificate for round $\le$ $v.round-1$ (if any)
                        }
                        \end{itemize}
			\item[] struct vote $vt$: \Comment{The struct of a vote}
			      \begin{itemize}[noitemsep]
			      	\item[] $vt.round$ - the round of $vt$
                        \item[] $vt.source$ - the party that broadcasts $vt$
                    {\color{blue}
                        \item[] $vt.strongEdges$ - leader vertices in $v.round-1$ that the party votes for; otherwise, $\bot$ 
                    }
			      \end{itemize}
                \item[] $\NoVote$ certificate $\NVC$ - an array of sets, each containing a quorum of $\NoVote$ messages
		\end{itemize}
            
		   \begin{minipage}{0.48\textwidth}
            \begin{algorithmic}
            \algrestore{bkbreak}
            
        \vspace{0.4em} 
             
        \Upon {receiving $\tuple{vt, r}_p$}
            \If {$vt.round=r$} 
            \State $votes[r]\gets votes[r]\cup \{vt\}$
            \If{$willProposeVertex=\True$}
            \State $VertexProposers[r]\gets VertexProposers[r]\cup \{p\}$
            \EndIf
            \State \textcolor{blue}{multicast $\tuple{vt, r}_p$}
            \EndIf
        \EndUpon
        
     \vspace{0.4em}   
             \Upon {$|votes[r]|+|DAG_i[r]| \geq n-f \land (\exists v' \in DAG_i[r] : v'.source = L_r\ \lor$ $\TC_r$ is received for $r \geq round)$}
        \color{blue}
        \State $rnd\gets r$ \Comment{round of the last main leader vertex delivered}
        \color{black}
        \If{$P_i=L_{r+1} \land \nexists v' \in DAG_i[r] : v'.source = L_r$}
            \State \textbf{wait until} receiving $\TC_{r}$
            \For{$r'=r-1$ down to $1$}
            \If{$\exists v'' \in DAG_i[r'] : v''.source = L_r'$}
                \color{blue}
                \State $rnd \gets r'$
                \color{black}
                \State \textbf{break}
            \Else 
                \State \textbf{wait until} receiving $\TC_{r'}$
            \EndIf
            \EndFor
        \EndIf
        \color{blue}
            \State advance\_round$(r+1,rnd)$
        \color{black}
        \EndUpon
             \Procedure {advance\_round}{$r,rnd$}
        \State $round\gets r$; $start\ timer$
        \color{blue}
        \For{$\ell \in \ML_{r-1} \land $}
            \If{$\not \exists v'\in DAG_i[r-1]:v'.source = \ell$}
            \State multicast $\tuple{\NoVote, \ell, r-1}_i$
            \EndIf
        \EndFor
        \If{$P_i=L_r$} 
        \State \textbf{wait until} $\exists v'\in DAG_i[rnd-1]: v'.source=\ell$ $\forall \ell \in \ML[:x]$ or $\NVC_{rnd-1}^{\ell'}$ is received, where $\ell'=\ML[x+1]$ \label{line: multi-leader wait}
        \State broadcast\_vertex($round$)
        \color{black}
        \ElsIf {$P_i$ intends to propose vertex in round $r$}
            \State broadcast\_vertex($round$)
        \Else
            \State multicast\_vote($round$)
        \EndIf
    \EndProcedure


         \algstore{bkbreak}
    \end{algorithmic}
    \end{minipage}
    \hfill
   \begin{minipage}{0.48\textwidth}
    \begin{algorithmic}
    \algrestore{bkbreak}

    \vspace{0.4em}  
    \Procedure{create\_new\_vertex}{$r$}
    \State $v.round \gets r$
    \State $v.source \gets \node{i}$
    \State $v.block \gets blocksToPropose$.dequeue()
    \If {$r>1$}
    \Upon {$|DAG_i[r-1]|\ge \operatorname{max}(|VertexProposers[r-2]|-f, 0)$} 
    \State $v.strongEdges \gets DAG_i[r-1]$
    \EndUpon
    \If{$\tuple{\Timeout,r-1}_i$ is sent}
    \State $v.strongEdges \gets v.strongEdges\ \backslash\ \{v':v'.source=L_{r-1}\}$
    \EndIf
    \EndIf
    \If{$\node{i} = L_r$ $\land$ $\not \exists v'\in DAG_i[r-1]:v'.source=L_{r-1}$}
    \For {$r'=r$ down to $1$}
    \If{$\not \exists v'' \in v.strongEdges \text{ s.t. } v''.source =L_{r'-1}$}
    \State $v.tc\gets v.tc\cup\{\TC_{r'-1}\}$
    \Else 
    \color{blue}
    \For{$\ell \in \ML_{r'-1}$} 
        \If{$\nexists v^* \in DAG_i[r'-1] : v^*.\text{source} = \ell$}
            \State $v.\mathit{nvc} \gets \NVC^{\ell}_{r'-1}$
            \State \textbf{break} 
        \Else
            \State $v.leaderEdges\gets v.leaderEdges \cup \{v^*\}$
        \EndIf
    \EndFor
    \State \textbf{break}
    \color{black}
    \EndIf
    \EndFor
    \EndIf
    \State \Call{set\_weak\_edges}{$v, r$}
    \State \Return $v$
    \EndProcedure
    
    \vspace{0.4em}
    
    \Procedure{create\_new\_vote}{$r$}
    \State $vt.round \gets r$
        \State $vt.source \gets \node{i}$
        \If {$\tuple{\Timeout,r-1}_i$ is sent}
        \color{blue}
        \State $vt.strongEdges\gets \bot$
        \Else
        \If {$r>1\land \exists v' \in DAG_i[r-1] : v'.source = L_{r-1}$} \label{line: multi-leader vote}
        \color{blue}
        \For{$\ell \in \ML_{r-1}$} 
        \If {$\exists v'' \in DAG_i[r-1] : v''.\text{source} = \ell$}
        \State $vt.strongEdges\gets vt.strongEdges\cup \{v''\}$
        \EndIf
        \EndFor
        \EndIf
        \EndIf
        \color{black}
        \State \Return $\tuple{vt, r}_i$
    \EndProcedure
    
    \end{algorithmic}
    \end{minipage}

\end{boxedminipage}
\caption{Multi-leader \name: DAG construction.}
\label{fig:multi-leader dag construction}
\end{figure*}

%% file: specs.tex
\section{Formal Specifications}
\label{sec:formal-specs}

In this section we provide a TLA+ specification of the single-leader \name protocol.
The model represents vertices, votes, and timeout messages as atomic posts to a global bulletin board. It abstracts Byzantine behavior, signatures and certificates (support is represented by sets of node identities), weak edges, transaction payloads, the implementation of reliable broadcast, real time, and the multi-leader extension.

Using TLC, we exhaustively checked three protocol-following nodes configured with $f=1$, quorums and blocking sets of size two, round-robin leaders, and three rounds. This is not an $n\geq 3f+1$ Byzantine configuration: because the model contains no deviating node, it exercises quorum intersection with the smallest configuration satisfying $n>2f$. The model checker terminated in 3 hours and 13 minutes using 32 workers on a 32-vCPU AMD EPYC virtual machine with 64 GB of memory. It generated 1,902,582,078 states, of which 18,938,251 were unique, and found no violations of either the type-correctness invariant or agreement. These checks provide confidence that the specification is executable and internally consistent, but they do not constitute a verification of Byzantine fault tolerance. The archived specification and model-checking configuration are available at \url{https://doi.org/10.5281/zenodo.21855535}.
\clearpage

\includepdf[pages=-]{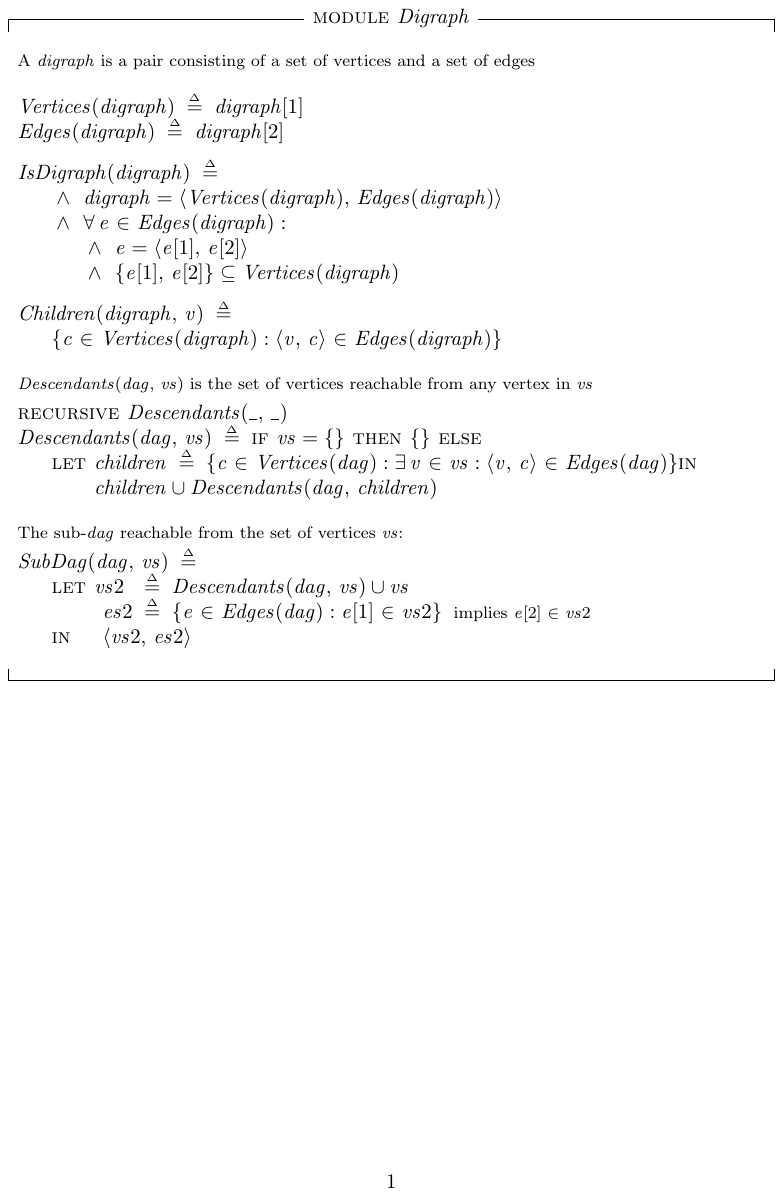}
\includepdf[pages=-]{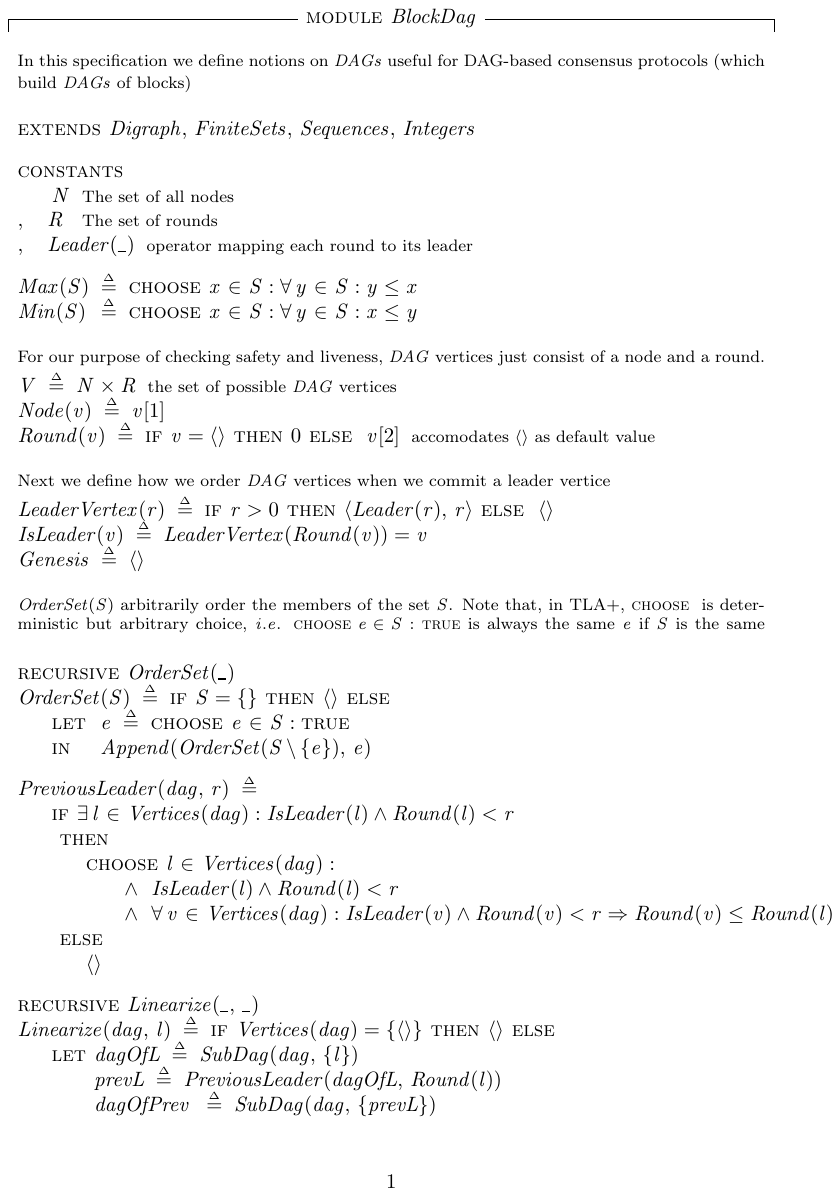}
\includepdf[pages=-]{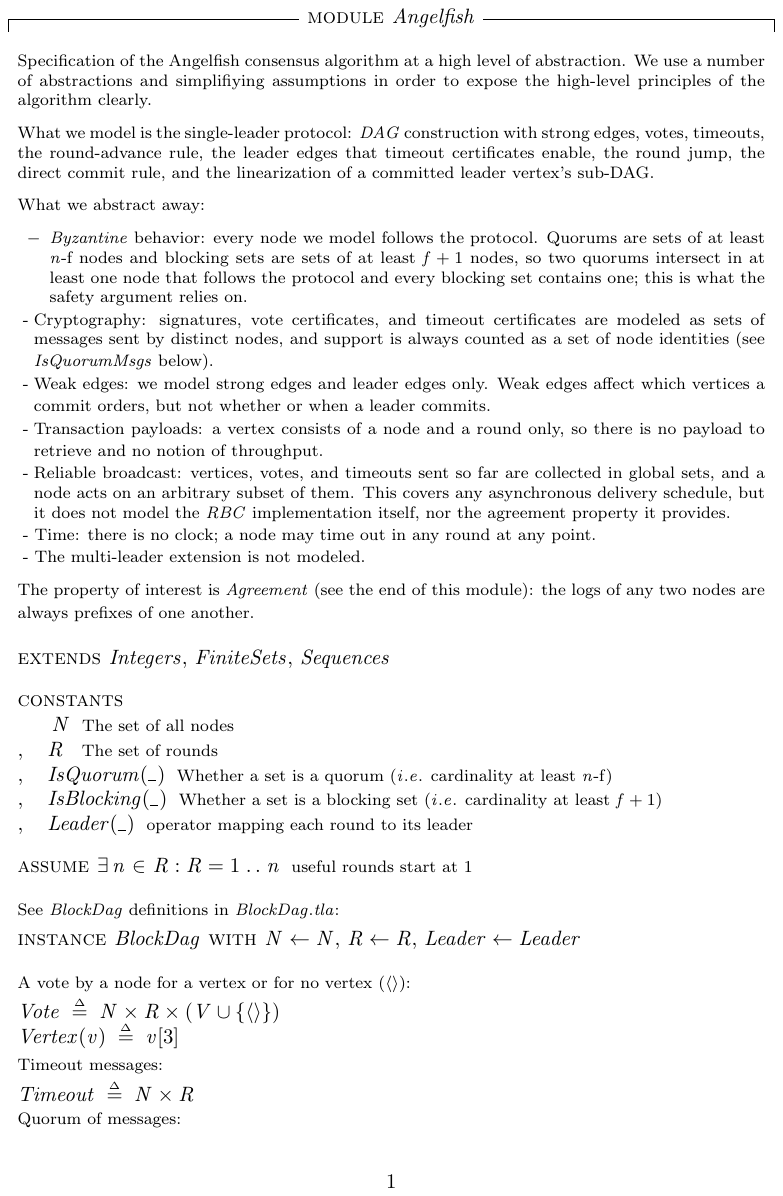}
\clearpage